\documentclass[12pt]{article}
\usepackage{latexsym}
\usepackage{hyperref}
\usepackage{epsfig,amssymb,euscript}
\usepackage{amsmath}
\usepackage{mathrsfs}
\newsavebox{\ns}
\newsavebox{\dbrane}
\newsavebox{\dbshort}

\def\be{\begin{equation}}
\def\ee{\end{equation}}
\def\bea{\begin{eqnarray}}
\def\eea{\end{eqnarray}}

\renewcommand{\theequation}{\arabic{section}.\arabic{equation}}

\def\theequation{\thesection.\arabic{equation}}
\def\be{\begin{equation}}
\def\ee{\end{equation}}
\def\ba{\begin{eqnarray}}
\def\ea{\end{eqnarray}}

\newcommand{\nn}{\nonumber}

\def\Dslash{\,\,{\raise.15ex\hbox{/}\mkern-12mu D}}
\def\Dbarslash{\,\,{\raise.15ex\hbox{/}\mkern-12mu {\bar D}}}
\def\delslash{\,\,{\raise.15ex\hbox{/}\mkern-9mu \partial}}
\def\delbarslash{\,\,{\raise.15ex\hbox{/}\mkern-9mu {\bar\partial}}}
\def\pslash{\,\,{\raise.15ex\hbox{/}\mkern-9mu p}}
\def\calDslash{\,\,{\raise.15ex\hbox{/}\mkern-12mu {\cal D}}}

\newcommand\R{\mathbb{R}}
\newcommand\Z{\mathbb{Z}}
\newcommand\F{\mathbb{F}}

\newcommand\D{\mathcal{D}}
\newcommand\cp{\mathbb{CP}}

\newcommand{\dd}{\mathrm{d}}

\newcommand{\vol}{\mathrm{vol}}

\newcommand{\Sym}{\mathrm{Sym}}

\def\NO{\nonumber}

\def\bea{\begin{eqnarray}}
\def\eea{\end{eqnarray}}

\def\beqx{\begin{displaymath}}
\def\eeqx{\end{displaymath}}

\newcommand{\bmat}{\left(\begin{array}}
\newcommand{\emat}{\end{array}\right)}

\def\b{\beta}

\def\d{\delta}

\def\f{\phi}
\def\g{\gamma}

\def\k{\kappa}
\def\l{\lambda}
\def\m{\mu}
\def\n{\nu}
\def\o{\omega}
    \def\om{\omega}
\def\p{\pi}

    \def\th{\theta}
\def\r{\rho}
\def\s{\sigma}
\def\t{\tau}

\def\z{\zeta}
\def\D{\Delta}
\def\F{\Phi}

    \def\Om{\Omega}

\def\X{\Xi}

\def\vf{\varphi}

\def\co{{\cal O}}
\def\cp{{\cal P}}

\def\cs{{\cal S}}

\def\bo{{\raise-.3ex\hbox{\large$\Box$}}}               
\def\pa{\partial}                                       
\def\face{{\raise.2ex\hbox{$\displaystyle \bigodot$}\mskip-2.2mu \llap {$\ddot
        \smile$}}}                                   
\def\>{\rangle}                                      
\def\<{\langle}                                      

\newcommand{\sub}[1]{\phantom{}_{(#1)}\phantom{}}    
\def\leftrightarrowfill{$\mathsurround=0pt \mathord\leftarrow \mkern-6mu
        \cleaders\hbox{$\mkern-2mu \mathord- \mkern-2mu$}\hfill
        \mkern-6mu \mathord\rightarrow$}        
\def\dvec#1{\vbox{\ialign{##\crcr
        \leftrightarrowfill\crcr\noalign{\kern-1pt\nointerlineskip}
        $\hfil\displaystyle{#1}\hfil$\crcr}}}           

\def\-{\hphantom{-}}

\begin{document}
\baselineskip=15.5pt
\pagestyle{plain}
\setcounter{page}{1}

\begin{titlepage}
\hfill CERN-PH-TH/2010-089

\vspace*{6mm}

\begin{center} 
\Large \bf The warped, resolved, deformed conifold\\[2mm]
gets flavoured
\end{center}

\vskip 11mm
\begin{center}
J\'er\^ome Gaillard$^a$\footnote{pyjg@swansea.ac.uk},
Dario Martelli$^b$\footnote{dario.martelli@kcl.ac.uk},
Carlos N\'u\~nez$^a$\footnote{c.nunez@swansea.ac.uk},
Ioannis Papadimitriou$^c$\footnote{Ioannis.Papadimitriou@cern.ch}
 \vskip 8mm

{\it $a$: Department of Physics, Swansea University\\
 Singleton Park, Swansea SA2 8PP, United Kingdom}
\vskip 7mm

{ \it $b$: Department of Mathematics, King's College, London\\
Strand, London WC2R 2LS, United Kingdom}

\vskip 7mm

\it{$c$: Department of Physics, CERN Theory Unit\\
1211 Geneva 23, Switzerland}

\vspace{0.3in}
\end{center}
\vskip 2.2cm
\begin{center}
{\bf Abstract}
\end{center}
We discuss  a simple transformation that allows to generate $SU(3)$ structure
solutions of Type  IIB supergravity with RR fluxes, starting from 
non-K\"ahler solutions of Type I supergravity.  The method may be applied also in the presence 
of supersymmetric source branes.
We apply this transformation to a solution describing fivebranes wrapped on the $S^2$ of the
resolved conifold with additional \emph{flavour} fivebrane sources.  The resulting solution is a 
generalisation of  the resolved deformed conifold solution of Butti \emph{et al} \cite{Butti:2004pk}
by the addition of D5 brane and D3 brane sources.  
We propose that  this solution  may be interpreted in terms of a combined 
effect of Higgsing and cascade of Seiberg  dualities in the dual field theory.

\vskip1truecm
\vspace{0.1in}
\end{titlepage}
\setcounter{footnote}{0}

\pagestyle{plain}
\setcounter{page}{1}
\newcounter{bean}
\baselineskip18pt
\tableofcontents

\section{Introduction}

The systematic study of supersymmetric geometries of String/M theory has provided new valuable tools
for addressing a number of problems in the context of the gauge/string 
duality as well as in string compactifications. 
Powerful techniques for studying supersymmetric backgrounds of various supergravities
are based on $G$-structures and generalised geometry. 
In this paper we will consider a class of supersymmetric backgrounds of Type  IIB supergravity
 characterised by an $SU(3)$ structure \cite{Grana:2005sn}, and we will
discuss a simple solution generating method applicable to these geometries.
 In particular, we will show that starting from a solution to the 
``torsional superstring'' equations of \cite{Strominger:1986uh}, 
a more general interpolating solution  may be generated, that includes the simple class of warped
Calabi-Yau solutions \cite{Dasgupta:1999ss} in a limit.  
In most cases,  this procedure is equivalent to the chain of U-dualities discussed in 
\cite{Maldacena:2009mw}, as was  also showed in \cite{Minasian:2009rn}.
However, exploiting the relation to generalised calibrations,
 this method can be applied as well to geometries which include the back-reaction of 
\emph{supersymmetric sources}.
In fact, we will apply the procedure to a supersymmetric solution describing 
$N_c$ D5 branes wrapped on the $S^2$ inside the resolved 
conifold, plus $N_f$ D5 branes sharing the $\R^{1,3}$ 
Minkowski directions and  infinitely 
extended along a transverse cylinder  \cite{Casero:2006pt}.    
We will study the case in which the $N_f$ 
sources are smeared over their transverse compact directions and we
can have  $N_f/N_c \sim {\cal O}(1)$. The motivations  and consequences of this smearing have been discussed in 
detail in \cite{Nunez:2010sf}.

The ``seed''  solution on which the solution generating 
technique  mentioned will be applied was discussed in 
\cite{HoyosBadajoz:2008fw}\footnote{More precisely, the solutions in \cite{HoyosBadajoz:2008fw} 
had different IR asympotics to the ones we will present in this paper.}. 
The large radius (UV) behaviour of this solution is such that the dilaton asymptotes to a constant
In section  \ref{newflavor} we will study the small radius (IR) asymptotics suitable for the purposes of this work, 
extending the analysis in \cite{HoyosBadajoz:2008fw,Casero:2007jj}. 
The solution we will discuss is singular at the origin of the radial direction
 and we may view this singularity as one that can be resolved, in a fashion similar to 
the Klebanov-Tseytlin solution \cite{Klebanov:2000nc}, where away from the singularity the 
solution captures the correct Physics. 
We will not resolve this singularity in the present paper, but we anticipate that 
one way to do this is to consider a profile for the smeared flavour branes that vanishes smoothly 
at the origin \cite{cgnr}.  The solution depends on two integers $N_c$ and $N_f$,  and 
we will fix the boundary conditions in the IR by requiring that setting $N_f=0$,
the smooth solution discussed in \cite{Maldacena:2009mw} is recovered.

The field theory proposed to be dual to the solutions
in \cite{Casero:2006pt,HoyosBadajoz:2008fw,Casero:2007jj}
is a deformation of ${\cal N}=1$ SQCD by coupling the quarks to the 
infinite tower of massive states (Kaluza-Klein modes) whose presence is due to the 
twisting of the original six-dimensional field theory. See \cite{Andrews:2006aw} for 
details about the twisting and the spectrum.

After applying the solution generating technique
 to the ``seed'' solution, we will find a new background with non-zero 
Ramond-Ramond (RR) and Neveu-Schwarz (NS) fields. In addition to the integers $N_c,N_f$, 
the solution depends on three continuous parameters. One of them is
the string coupling at infinity.  
Then there is a parameter we denote $c$, which is 
related to the size of the $S^2$ as measured from infinity \cite{Maldacena:2009mw}, 
hence to the amount of resolution in the geometry. 
In the unflavoured case this parameter was related to the VEV of baryonic operators 
in the baryonic branch of the Klebanov-Strassler theory \cite{Klebanov:2000hb}. 
Finally, we have a parameter introduced by the transformation\footnote{As we will explain 
later, this parameter corresponds to the boost parameter in \cite{Maldacena:2009mw}.}.
The solution may be viewed as a ``flavoured'' version of the resolved deformed conifold
solution \cite{Butti:2004pk}, thus justifying our title. 
Sending to zero the resolution parameter  ($c \to \infty$), and $N_f$,
we will  obtain a solution closely related to the Klebanov-Strassler solution.

We will then discuss the dual field theory interpretation of our new solutions. 
We begin with the simpler solution mentioned above: in this case we will argue that the 
presence of smeared source D3 branes 
may be interpreted as a change of ranks of the gauge theory due to  \emph{Higgsing}, 
that in our case will take place at all energy scales.  This is inspired by the ideas
discussed in \cite{Aharony:2000pp}.  
More precisely, we will propose a Klebanov-Strassler type 
quiver theory $SU(N_c+n+n_f)\times SU(n+n_f)$ in the mesonic branch,
where the change of ranks is due to a combination\footnote{We are indebted to 
Ofer Aharony for crucial comments  that helped us sharpening this 
proposal.}  of Seiberg dualities (running of $n$) and of Higgsing (running of $n_f$). 

The field theory interpretation of the general case when we
restore a non-zero value of $N_f$ is more complicated. We will make some general comments and then 
we will analyse two kinds of field theory scenarios.  In both cases we will consider 
a quiver gauge theory with two nodes, each with different ranks of the gauge groups,  
and we will assume that there are bi-fundamental fields 
interacting with a quartic superpotential,  precisely as in the Klebanov-Strassler theory.  
In section \ref{sce1}, we will analyse a quiver in which we have added \emph{explicit flavours}, 
\emph{i.e.} fields transforming in fundamental representations of the gauge groups. 
In this case the global flavour  
symmetry is manifest along the flow from the UV to the IR. However, in this picture
a precise understanding of the interplay of Seiberg dualities and Higgsing 
appears problematic, as we will explain.  
 In section \ref{sce2}  we will discuss the possibility that the 
flavour symmetry  may develop dynamically after Higgsing and the cascade take place. 
As we will see,  in this case we can obtain a rather detailed matching between field theory and gravity computations.

The rest of the paper is organized as follows: in section \ref{gsfisu3} 
we will discuss a solution generating method applicable to supersymmetric 
 Type IIB geometries  characterised by an $SU(3)$ structure.  We will also explain how to incorporate
supersymmetric  sources (flavour branes).  In section \ref{rtfwd5s} we 
will give details of the new solutions. First we will extend previous 
studies on fivebrane solutions with asymptotically constant dilaton, 
 and then we will construct explicitly the new flavoured resolved deformed conifold 
 solution.  In section \ref{comme} we will discuss possible  
dual field theory interpretations of these backgrounds.  The two scenarios 
alluded to above will be discussed in detail.  Section \ref{thez2}
contains a discussion of a $\Z_2$ symmetry, and its breaking,
in gravity and in the field theory.
Finally, we summarise our findings in section \ref{sectiondiscussionzz}. 
The appendices include the Type IIB equations of motion and  various
technical calculations that we will quote in the paper.

\section{Generating solutions from  $SU(3)$ structures}
\label{gsfisu3}

We start this section by presenting the BPS equations of an $SU(3)$ structure
background, derived  from the  general set-up in \cite{Grana:2005sn}.
While the authors of \cite{Grana:2005sn}  work with pure spinors, in a 
particular case, their results can be formulated 
in terms of the two differential forms characterising  the
$SU(3)$ structure. First, let us state precisely the ansatz and conventions.
We work with Type IIB supergravity in Einstein frame and consider a
ten-dimensional space as a warped product of a
four-dimensional Minkowski space and a six-dimensional
space equipped with an $SU(3)$ structure, using the metric
\be\label{metric-rotation}
	\dd s^2 = e^{2\Delta} \big[ \dd x_{1,3}^2 + \dd s^2_6 \big].
\ee
We also have several fluxes: the RR forms $F_{(1)}$, $F_{(3)}$, $F_{(5)}$ and the NS three-form $H$. 
These  have components only in the
internal six-dimensional space, except for $F_{(5)}$ that is self-dual.
Generically, we can write
\be
	\begin{aligned}
		F_{(5)} &= e^{4\Delta + \Phi} (1+ *_{10}) \vol_{(4)} \wedge f ,\\
		H &= \dd B, \\
	\end{aligned}
\ee
where $f$ is a one-form. The general Type IIB supersymmetry 
conditions for these geometries 
were  derived in \cite{Grana:2005sn,Martucci:2005ht} as equations  
for the two pure spinors (multi-forms)  $\Psi_1,\Psi_2$ and read
\be
	\begin{aligned}
		e^{-2\Delta+\Phi/2}\! (\dd - H \wedge)\! \left[ e^{2\Delta-\Phi/2} \Psi_1 \right]\! &=\! \dd\! \left(\!\Delta + \frac{\Phi}{4}\! \right)\! \wedge\! \bar{\Psi}_1 +\! \frac{i e^{\Delta + 5\Phi/4}}{8}\! \left[ f -\! *_6 F_{(3)}\! + e^{4\Delta+\Phi}\! *_6 F_{(1)} \right], \\
		(\dd - H \wedge)\! \left[ e^{2\Delta-\Phi/2} \Psi_2 \right]\! &= 0.
	\end{aligned}
\ee
We then specialise these to the case of $SU(3)$ structure. This means that the two pure spinors  take the form
\be
\Psi_1 = - \frac{e^{i \zeta}}{8} e^{\Delta + \Phi/4} \left( 1 - i e^{2\Delta + \Phi/2} J - \frac{1}{2} e^{4\Delta + \Phi} J \wedge J
\right) ,\;\;\; \Psi_2 = - \frac{e^{4\Delta + \Phi}}{8} \Omega~.
\ee
Here $J$ is the (would-be) K\"ahler two-form  and $\Omega$ is the
holomorphic three-form. Together, these define  an 
$SU(3)$ structure on  the six-dimensional geometry.
In addition, we have a function  $\zeta$ arising as a phase in the pure spinor $\Psi_1$. 
For this reason, the  $SU(3)$ structure is referred to as 
interpolating\footnote{See  \cite{Martelli:2003ki,Frey:2003sd} for earlier work on interpolating geometries.}. 

Equating terms involving forms of the same degree
we obtain  the BPS equations of the system written as
\be \label{eq:genBPS}
	\begin{aligned}
		\dd \left( e^{6\Delta + \Phi/2} \Omega \right) &= 0 \\
		\dd \left( e^{8\Delta} J \wedge J \right) &= 0 \\
		\dd \left( e^{2\Delta -\Phi/2} \cos \zeta \right) &= 0 \\
		- e^{-4\Delta - \Phi} \dd \left( e^{4\Delta} \sin \zeta \right) &= f \\
		- e^{\Phi} \cos \zeta *_6 F_{(3)} - e^{2\Delta + 3\Phi/2} \sin \zeta \dd \left(e^{-\Phi} \sin \zeta \right) \wedge J &= e^{-2\Delta - \Phi /2} \dd \left( e^{4\Delta + \Phi} J  \right) \\
		- \sin \zeta e^{\Phi} *_6 F_{(3)} + \cos \zeta e^{2\Delta + 3\Phi/2} \dd \left(e^{-\Phi} \sin \zeta \right) \wedge J &= H \\
		-\frac{1}{2} \dd \left(e^{-\Phi} \sin \zeta \right) \wedge J \wedge J &= *_6 F_{(1)}
	\end{aligned}
\ee
Manipulating these equations a little more one can show that
\be
	H = \dd \left(\tan \zeta e^{2\Delta + \Phi/2} J \right)~~~\to~~ ~	B
= \tan \zeta e^{2\Delta + \Phi/2} J~,
\ee
thus the (non-closed part of the) $B$ field is determined by the $SU(3)$ structure.
Notice that the first  equation in (\ref{eq:genBPS}) implies that the geometry is \emph{complex}, in the usual sense, as opposed to the general case discussed in \cite{Grana:2005sn,Martucci:2005ht}. This gives a useful 
characterisation of the geometries we are interested in.
These equations were also derived  in \cite{Minasian:2009rn},
which discussed first the results presented in this section.

In the rest of this paper, we will impose that $F_{(1)} = 0$.
The last equation in  \eqref{eq:genBPS} then implies that $\dd \left(e^{-\Phi} \sin
\zeta \right) = 0 $ and the system simplifies further, reducing to 
\be \label{eq:rotatedBPS}
	\begin{aligned}
\dd \left( e^{6\Delta + \Phi/2} \Omega \right) &= 0,\;\;~~~~\dd \left( e^{2\Delta -\Phi/2} \cos \zeta \right) = 0,\;\;~~~~
\dd \left( e^{8\Delta} J \wedge J \right) = 0, \\
- e^{\Phi} \cos \zeta *_6 F_{(3)} &= e^{-2\Delta - \Phi /2} \dd \left( e^{4\Delta + \Phi} J  \right),\;\;~~~~ H=- \sin \zeta e^{\Phi} *_6 F_{(3)}, \\
- e^{-4\Delta - \Phi} \dd \left( e^{4\Delta} \sin \zeta \right) &= f.
\end{aligned}
\ee
It is instructive to specialise
 the system 
 \eqref{eq:rotatedBPS} to the case  $\zeta = 0$: 
\be \label{eq:specBPS}
	\begin{aligned}
		\dd \left( e^{6\Delta + \Phi/2} \Omega \right) &= 0,\;\;~~~~\dd \left( e^{2\Delta -\Phi/2} \right) = 0, \;\;~~~~
\dd \left( e^{8\Delta} J \wedge J \right) = 0 \\
- e^{\Phi} *_6 F_{(3)} &= e^{-2\Delta - \Phi /2} \dd \left( e^{4\Delta + \Phi} J  \right) \\
		f &= 0,\;\;\;~~ H=0,\;\;\;~~  F_{(1)} =0. \\
\end{aligned}
\ee
The only non-zero flux is then $F_{(3)}$ and 
the BPS system describes a configuration of D5 branes. These are simply the S-dual version of the 
``torsional superstring'' equations of  \cite{Strominger:1986uh,Hull:1986kz} and 
 they were written in this form in  \cite{Gauntlett:2001ur,GMW}. 
A notable solution to these equations was discussed in 
 \cite{Maldacena:2000yy}.

We will now show that from a solution of the  system \eqref{eq:specBPS}
one can generate a solution of the more complicated system \eqref{eq:rotatedBPS} for a non-vanishing
$\zeta$. This is then a simple solution generating technique. 
We will sometimes refer to this procedure as 
$\it{rotation}$\footnote{There are 
different motivations for this: firstly, it is a rotation in the space of 
Killing spinors, parameterised by $\zeta$. Secondly, in the particular case discussed in \cite{Maldacena:2009mw} 
this corresponds to an actual rotation of NS5 branes in a T-dual Type IIA brane picture.}. 
Precisely, we have that if one defines
\be 
\label{eq:rotation_formula}
	\begin{aligned}
		\Phi &= \Phi^{(0)} \\
		e^{2\Delta} &= \frac{1}{\cos \zeta} e^{2\Delta^{(0)}} = \frac{\kappa_1}{\cos \z} e^{\F/2} \\
		\Omega &= \left( \frac{\cos \zeta}{\kappa_1} \right)^3 \Omega^{(0)} \\
		J &= \left( \frac{\cos \zeta}{\kappa_1} \right)^2 J^{(0)} \\
		F_{(3)} &= \frac{1}{\kappa_1^2} F_{(3)}^{(0)} \\
		F_{(5)} &= - (1+*_{10}) \dd x^0 \wedge \dd x^1 \wedge \dd x^2 \wedge \dd x^3 \wedge \dd \left( \frac{\sin \zeta}{\cos^2 \zeta} e^{4 \Delta^{(0)}} \right)
	\end{aligned}
\ee
where the quantities with a $^{(0)}$ obey the equations in \eqref{eq:specBPS}, then the (new)
 quantities on the left-hand side obey the equations in \eqref{eq:rotatedBPS}. $\kappa_1$ is here an integration constant. 
We require that the condition $F_{(1)} = 0$ is preserved, which implies 
that $\dd \left( e^{-\Phi} 
\sin \zeta \right) =0$. We can then solve this equation obtaining 
\be
	\sin \zeta = \kappa_2 e^{\Phi}~,
\label{importanteq}
\ee
where $\kappa_2$ is another integration constant. This formula requires the dilaton 
to be bounded from above at any position in space. To summarise, let us 
write the background 
after the ``rotation''
 in terms of the 
initial one (in Einstein frame),
\be\label{rotated-background}
	\begin{aligned}
		\dd s^2 &= e^{-\F/2} \left[ h^{-1/2} \dd x_{1,3}^2 + e^{2\F}h^{1/2} \dd s_6^{(0)2} \right] \\
		F_{(3)} &= \frac{1}{\k_1}e^{-2\F}\ast_6 \dd \left( e^{2\F}J^{(0)} \right) \\
		B &= \frac{\k_2}{\k_1}e^{2\F}J^{(0)} \\
		F_{(5)} &= -\k_2(1+\ast_{10}) \vol_{(4)} \wedge \dd h^{-1}
	\end{aligned}
\ee
where
\be\label{h}
	h=\frac{1}{\k_1^2}\left(e^{-2\F}-\k_2^2\right)~.
\ee

Any solution of the system \eqref{eq:genBPS}, supplemented by the Bianchi identities for the fluxes, is a solution of the equations 
of motion of Type  IIB supergravity \cite{Grana:2005sn,Martucci:2005ht}. 
 One can then show that imposing the Bianchi identities for the simplified 
(seed) system \eqref{eq:specBPS}, implies also the Bianchi identities, and 
hence the full equations of motion, of the more complicated system. 
Thus, starting from a solution of the system \eqref{eq:specBPS} (with a 
bounded dilaton), one can generate a solution of the system \eqref{eq:rotatedBPS} using the formulas in  \eqref{eq:rotation_formula}.  This result was discussed also in \cite{Minasian:2009rn}.

\subsection{Adding D5 brane sources}

In this subsection we will show how the generating technique discussed above may be applied 
also to supersymmetric solutions for a combined system of supergravity plus smeared sources. 
The key observations are the following. Firstly, when the sources are smeared in a supersymmetric way, 
the  D-brane action can be written in terms of generalised calibrations 
and their net effect is captured by simple modifications of the Bianchi identities for the fluxes. 
In particular, the non-closed part of the fluxes is identified with the so-called smearing form -- see \cite{Gaillard:2008wt}.  
The supersymmetry equations, however, do not change in form. Then by using the results of \cite{Koerber:2007hd}, 
the computation of the previous subsection can be applied to the case with sources.
The interest of including the back-reaction of such 
explicit branes is that these may be interpreted as flavours in the context of the gauge/gravity duality.

We  consider the case of D5 brane sources. One then needs to study the combined action of 
Type IIB supergravity with the DBI and WZ terms for the source branes. 
Using the $SU(3)$ structure calibration conditions, 
the combined action can be written as \cite{Gaillard:2008wt}
\be \label{eq:flavoured_action}
	S = S_{IIB} - \int \Big(e^{4\Delta + \Phi/2} \vol_{(4)} \wedge \big( \cos \zeta e^{2\Delta} J + \sin \zeta e^{-\Phi/2} B \big) - C_{(6)} + C_{(4)} \wedge B \Big) \wedge \Xi_{(4)}
\ee
where $S_{IIB}$ is the action of Type IIB supergravity. 
Here $\Xi_{(4)}$ is the smearing form, characterising the distribution 
of sources. It is proportional to the number 
of flavour branes $N_f$ and has no components along  the Minkowski 
directions. $C_{(4)}$ and $C_{(6)}$ are defined via
\be \label{eq:potentials}
	\begin{aligned}
		F_{(5)} &= \dd C_{(4)} + B \wedge F_{(3)}, \\
		F_{(7)} &= -e^{\Phi} *_{10} F_{(3)} = \dd C_{(6)} + B \wedge F_{(5)}.
	\end{aligned}
\ee
Using the results of \cite{Koerber:2007hd}, 
we know that the  addition of  sources, even when smeared, does not modify
the form of the BPS system  \eqref{eq:genBPS} but only the Bianchi identities. 
These now read
\be
	\begin{aligned}
		\dd F_{(3)} &= \Xi_{(4)} ,\\
		\dd F_{(5)} &= H \wedge F_{(3)} + B \wedge \Xi_{(4)}.
	\end{aligned}
\label{f5bianchi}
\ee
As shown in \cite{Koerber:2007hd}, if one imposes the Bianchi 
identities, every solution  of the BPS
system is a solution of the equations of motion
coming from \eqref{eq:flavoured_action}.
As previously described, one can generate a
solution to the equations of motion of the action \eqref{eq:flavoured_action}
from the case $\zeta = 0$. Setting $\zeta = 0$ in
\eqref{eq:flavoured_action} and using the fact that $B = 0$
we find
\be \label{eq:spec_flavoured_action}
	S_{\text{D5 sources}}  =  - \int \Big(e^{6\Delta + \Phi/2} \vol_{(4)} \wedge J - C_{(6)} \Big) \wedge \Xi_{(4)}.
\ee
This  is the action for supersymmetric D5 brane sources in a background characterised by the equations \eqref{eq:specBPS}.

\subsection{The limit of D3 brane sources}

\label{limitsection}

The limit $\zeta \to \pi/2$, in which the supergravity background goes over to the 
warped Calabi-Yau geometry, is slightly more subtle \cite{Maldacena:2009mw}.
Here we will determine
how the action for the source branes behaves in this limit, and we will find 
that indeed the limiting action
corresponds to smeared source D3 branes, with a particular smearing form arising in the limit. 
Considering  the action \eqref{eq:flavoured_action}
as generated from the $\zeta = 0$ case \eqref{eq:spec_flavoured_action}, 
we can work out the dependence on $\zeta$ of every
quantity from eqs. \eqref{eq:rotatedBPS} and \eqref{eq:rotation_formula}.
Recalling that the RR potentials are defined using \eqref{eq:potentials}, 
we have
\be
	\begin{aligned}
		e^{2\Delta} &= \frac{1}{\cos \zeta} e^{2 \Delta^{(0)}} = \frac{\k_1}{\cos \zeta} e^{\F/2} \\
		e^{\Phi} &= \frac{1}{\kappa_2} \sin \zeta \\
		J &= \frac{\cos^2 \zeta}{\k_1^2} J^{(0)} \\
		B &= \frac{1}{\kappa_1^2 \sqrt{\kappa_2}} \sin^{3/2} \zeta e^{2\Delta^{(0)}} J^{(0)} \\
		C_{(6)} &= \frac{1}{\kappa_1^2 \sqrt{\kappa_2}} \sqrt{\sin \zeta} e^{6\Delta^{(0)}} \vol_{(4)} \wedge J^{(0)} \\
		C_{(4)} &= - \frac{\sin \zeta}{\cos^2 \zeta} e^{4 \Delta^{(0)}} \vol_{(4)}
	\end{aligned}
\ee
Looking at the dependence on $\cos \zeta$, the action for the sources can be written as
\be
	S_{\text{sources}} = - \int \frac{1}{\cos^2 \zeta} \Big(\big[\cos^2 \zeta \sin \zeta e^{4\Delta} \vol_{(4)} \wedge B + \cos^2 \zeta C_{(4)} \wedge B \big] + O(\cos^2 \zeta) \Big) \wedge \Xi_{(4)}.
\ee
In this formula, the quantity in square brackets does not
scale with $\cos \zeta$, implying 
that this  is approaching a finite non-zero value when
$\zeta$ goes to $\pi/2$, and additional terms in $O(\cos^2 \zeta)$
go to zero in the limit. However, there is an overall
factor $\cos^{-2} \zeta$. Therefore, if we want the action to be finite in the
limit $\cos \zeta \rightarrow 0$, then we need to
scale $\Xi_{(4)}$ accordingly. However, $\Xi_{(4)} = N_f \omega_{(4)}$
where $\omega_{(4)}$ does not depend on $\zeta$.
We then conclude that we need to impose the following condition:
\be
	\frac{N_f}{\cos^2 \zeta} \rightarrow \mathrm{constant}~~~\text{  when  }~~~ \zeta \rightarrow \frac{\pi}{2}.
\ee
In this case the limit of the D5 brane source action is
\be
	S_{\text{sources}} ~\rightarrow~ S_{\text{D3 sources}}= - \int \Big(e^{4\Delta^{(0)}} \vol_{(4)} + \tilde{C}_{(4)} \Big) \wedge \tilde{\Xi}_{(6)}
\ee
where the tilded quantities correspond to the limit
of the untilded ones. We have defined $\tilde{\Xi}_{(6)}$ as
\be
	B \wedge \Xi_{(4)} \rightarrow \tilde{\Xi}_{(6)} ~~~~\text{  when  } ~~~~\zeta \rightarrow \frac{\pi}{2}.
\ee
In the limiting case we can then identify the action as the smearing of supersymmetric D3 branes with smearing form $\tilde{\Xi}_{(6)}$.

\section{Adding D3 branes to  the flavoured D5 brane solution}
\label{rtfwd5s}
\setcounter{equation}{0}

Here we apply the procedure discussed in the previous section to a solution
 representing D5 branes wrapped on the $S^2$ of the resolved conifold, with the addition of explicit smeared D5 brane sources. The resulting Type IIB solution with D3 brane charge and $B$ field 
will be a ``flavoured'' version of the warped resolved deformed conifold
solution originally derived in \cite{Butti:2004pk}.
We will postpone a field theory interpretation of this new background until 
section \ref{comme}.

\subsection{D5 branes on the resolved conifold with flavour D5 branes}
\label{d5section}

The setup corresponding to D5 branes wrapped on the $S^2$ of the resolved conifold, with addition of smeared
D5 sources was described in 
\cite{Casero:2006pt,HoyosBadajoz:2008fw,Casero:2007jj}. 
The metric in the Einstein frame takes the form
\be\label{metric-rotated}
\dd s^2=e^{2\D}\left[ \dd x^2_{1,3}+e^{\F-4\D}\dd s_6^2\right],
\ee
where here the internal metric $\dd s_6^2$ does not change under the ``rotation'' procedure. 
As we have seen in the previous
section, the un-rotated metric is obtained by 
setting $e^{2\Delta} =  e^{2\Delta^{(0)}} =\kappa_1 e^{\F/2}$ in (\ref{metric-rotated}). 
The solution can then be completely described\footnote{In Appendix \ref{mmvar} we have
 written this solution in the variables used in  \cite{Maldacena:2009mw}.}
 in terms of the 
sechsbeins $e^a$, $a=1,\ldots,6$, parameterising the
internal metric $\dd s_6^2$, i.e. $\dd s_6^2=\d_{ab}e^ae^b$.
In the notation of \cite{Casero:2006pt} the sechsbeins are 
\bea
&&e^\r=e^{k}\dd\r,\quad e^\th=e^{q}\om_1,\quad e^\vf=e^{q}\om_2,\NO\\
&&e^1=\frac12e^{g}\left(\tilde{\om}_1+a\om_1\right),\quad e^2=\frac12 e^{g}\left(\tilde{\om}_2-a\om_2\right),
\quad e^3=\frac12e^{k}(\tilde{\om}_3+\om_3),
\eea
where the one-forms $\om_i$, $\tilde{\om}_i$, $i=1,2,3$, are defined as
\be
\begin{array}{ll}
\om_1=\dd\th, & \tilde{\om}_1=\cos\psi \dd\tilde{\th}+\sin\psi \sin\tilde{\th}\dd\tilde{\vf}, \\
\om_2=\sin\th \dd\vf, & \tilde{\om}_2=-\sin\psi \dd\tilde{\th}+\cos\psi\sin\tilde{\th} \dd\tilde{\vf},\\
\om_3=\cos\th \dd\vf, & \tilde{\om}_3=\dd\psi+\cos\tilde{\th}\dd\tilde{\vf}.
\end{array}
\ee
Moreover, the RR three-form is given by
\bea
F\sub{3}&=&-2 N_c e^{-2g -k}e^1\wedge e^2\wedge e^3
+\frac{N_c}{2}(a^2 -2 a b +1 -\frac{N_f}{N_c}) e^{ -2q -k}e^\th\wedge e^\vf\wedge e^3\NO\\
&&+N_c(b-a) e^{ -g -q-k}(e^1\wedge e^\vf+e^2\wedge e^\th)\wedge e^3\NO\\
&&+\frac{N_c}{2}b'e^{ -g-q-k}e^\r\wedge(-e^\th\wedge e^1+e^\vf\wedge e^2).
\label{3-form}
\eea
As was shown in \cite{HoyosBadajoz:2008fw}, there is a set of variables that decouples the BPS
equations described in the previous section for this ansatz and leads to a single second order ordinary differential
equation, whose solution completely determines the supergravity background. See also Appendix \ref{mmvar}.
This set of variables is introduced defining
\bea
&e^{2q}=\frac14\left(\frac{P^2-Q^2}{P\cosh\t-Q}\right),
\quad e^{2g}=P\cosh\t-Q,\quad e^{2k}=4Y,\NO\\
&a=\frac{P\sinh\t}{P\cosh\t-Q},\quad b=\frac{\s}{N_c}.
\label{changevariables}
\eea
Solving the resulting set of decoupled BPS equations one then finds
\bea\label{BPS-sols}
\begin{aligned}
\sinh\t =&\frac{1}{\sinh(2(\r-\r_o))},\\
Q=&\left(Q_o+\frac{2N_c-N_f}{2}\right)\cosh\t+\frac{2N_c-N_f}{2}\left(2\r\cosh\t-1\right),\\
\s =&\tanh\t\left(Q+\frac{2N_c-N_f}{2}\right),\\
e^{4(\F-\F_o)}=&\frac{\cosh^2(2\r_o)}{(P^2-Q^2)Y\sinh^2\t},\\
Y=&\frac18(P'+N_f),
\end{aligned}
\eea
while the only remaining unknown function, $P(\r)$, is determined by the equation
\be\label{master}
P''+(P'+N_f)\left(\frac{P'+Q'+2N_f}{P-Q}+\frac{P'-Q'+2N_f}{P+Q}-4\cosh\t\right)=0.
\ee
Here $\r_o$, $Q_o$ and $\F_o$ are constants of integration 
and we set $Q_o = -N_c + N_f/2$.
The $SU(3)$ structure for this class of backgrounds is specified by a 
(would-be) K\"ahler
form $J$ and a holomorphic three-form $\Om$, which may be written 
explicitly 
as follows (cf. \cite{Maldacena:2009mw} for the unflavored case)
\bea
\label{su3-forms}
J^{(0)}&=&e^\r\wedge e^3+e^\th\wedge\left(-\cos\m e^\vf+\sin\m e^2\right)
+e^1\wedge \left(-\sin\m e^\vf-\cos\m e^2\right),\\
\Om^{(0)}&=&\left(e^\r+ie^3\right)\wedge[e^\th+i\left(-\cos\m e^\vf
+\sin\m e^2\right)]\wedge [e^1+i\left(-\sin\m e^\vf-\cos\m e^2\right)],\NO
\eea
where the angle $0<\m<\p/2$ corresponds to a rotation in the $e^\vf-e^2$ plane and is given by
\be\label{angle-constraint}
\cos\m=\frac{P-\cosh\t Q}{P\cosh\t-Q}.
\ee
The $SU(3)$ structure for the transformed solution is now obtained simply from (\ref{su3-forms}) via the rescalings (\ref{eq:rotation_formula}).

\begin{flushleft}
{\em Charge quantisation}
\end{flushleft}

Given a solution of (\ref{master}) one immediately obtains the full string background via the above relations. 
In particular, the new background is obtained as in (\ref{rotated-background}) and it depends 
on the parameters $N_c$ and $N_f$, which in the case $\z=0$,
can be interpreted respectively as the number of colour and flavour D5 branes. 
However, this interpretation should be reconsidered for the transformed backgrounds.
Notice that in the presence of sources one should be careful with 
the D5 charge quantisation condition for the original background. In particular, since $F_3$ is not closed, its integral 
over the three-cycle at infinity will depend on the representative submanifold, hence it cannot be quantised.  
We therefore \emph{define} the number of colour D5 branes by integrating over the three-cycle $F_3$ evaluated at $N_f=0$.
Since the latter is closed, this definition makes sense and we have
\be
\frac{1}{2\k_{10}}\int_{S^3}F_3|_{N_f=0}=N_cT_5,
\ee
where $S^3$ is any representative of the unique three-cycle at infinity. 
For the transformed background this should be modified to
\be
\frac{1}{2\k_{10}}\int_{S^3}F_3|_{N_f=0}=\widetilde{N}_cT_5=\frac{N_c}{\k_1}T_5 \in \mathbb{N}.
\ee
We also redefine the number of flavours using
\be
\dd F_3=\frac{\widetilde{N}_f}{4}\sin\th\sin\tilde{\th}d\th\wedge \dd\vf\wedge \dd\tilde{\th}\wedge \dd\tilde{\vf}=
 \frac{N_f}{4\k_1}\sin\th\sin\tilde{\th}\dd\th\wedge \dd\vf\wedge \dd\tilde{\th}\wedge \dd\tilde{\vf}.
\ee
Noticing that $P$, $Q$ and $J^{(0)}$ are homogeneous 
of $N_c$ and $N_f$ of degree one,
it follows that the  rotated background, in \emph{string frame}, becomes
\bea
&&\dd s_{str}^2=h^{-1/2}\dd x_{1,3}^2+e^{2\F}h^{1/2}\k_1 \dd s_6^2(\widetilde{N}_c,\widetilde{N}_f)\NO\\
&&F_3=e^{-2\F}\ast_6 \dd(e^{2\F}J^{(0)}(\widetilde{N}_c,\widetilde{N}_f)),\NO\\
&&B=\k_2 e^{2\F}J^{(0)}(\widetilde{N}_c,\widetilde{N}_f),\NO\\
&&F_5=-\k_2(1+\ast_{10})\dd h^{-1}\wedge \vol_{(4)}.
\eea
Using the expression (\ref{h}) for $h$,  it is clear that the constant 
$\k_1$ can be absorbed into the   
rescaled charges $\widetilde{N}_c$ and $\widetilde{N}_f$. Namely, defining
\be
\hat{h}=e^{-2\F}-\k_2^2,
\label{newwarp}
\ee
and absorbing $\k_1$ by a trivial rescaling of the worldvolume coordinates $x^i\to \k_1^{-1} x^i$, we have 
\bea\label{rotated-solution}
&&\dd s_{str}^2=\hat{h}^{-1/2}\dd x_{1,3}^2+e^{2\F}\hat{h}^{1/2}\dd s_6^2(\widetilde{N}_c,\widetilde{N}_f)\NO\\
&&F_3=e^{-2\F}\ast_6\dd(e^{2\F}J^{(0)}(\widetilde{N}_c,\widetilde{N}_f)),\NO\\
&&B=\k_2 e^{2\F}J^{(0)}(\widetilde{N}_c,\widetilde{N}_f),\NO\\
&&F_5=-\k_2(1+\ast_{10})\dd \hat{h}^{-1}\wedge \vol_{(4)}.
\eea
It follows that the effect of the 
rotation described in the previous section is simply 
the introduction of the parameter $\k_2$, 
with $0\leq \k_2 < \max\{e^{-\F}\}$. For $\k_2=0$ we recover the original background. To 
make contact with the discussion in \cite{Maldacena:2009mw}, we may
 parameterise $\k_2$ as
\be
\k_2=e^{-\F_\infty}\tanh\b,
\label{kappa2xx}
\ee
where $\F_\infty$ is the asymptotic value of the dilaton.  In \cite{Maldacena:2009mw} this transformation was derived as a simple chain of U-dualities, and the constant $\beta$ 
arose as a boost parameter in eleven dimensions. However, the derivation presented here (see also \cite{Minasian:2009rn})
may be readily applied to cases with sources.

\subsection{The flavoured  resolved deformed conifold}

We will now present a deformation 
of the  solution of Butti \emph{et al}  \cite{Butti:2004pk},  describing
the baryonic branch of the Klebanov-Strassler  theory \cite{Klebanov:2000hb},
induced by the back-reaction of source D5 branes. 
We will start by first  reviewing some of the material in  \cite{Maldacena:2009mw}.

\subsubsection{Review of the unflavoured solution}

Before we present the flavoured solution, let us recall the unflavoured 
solution \cite{Butti:2004pk,Casero:2006pt,HoyosBadajoz:2008fw,Maldacena:2009mw}. This solution is obtained by setting 
$\widetilde{N}_f=0$ in (\ref{rotated-solution}), $Q_o=-\widetilde{N}_c$, $\r_o=0$ and picking a 
specific solution of the differential equation (\ref{master}) for $P$. The solution is only known numerically, 
but one can easily determine its IR and UV asymptotic forms, which are specified in terms of two arbitrary 
constants, $h_1$ and $c$:\footnote{These constants are related to the parameters $\g$, $t_{\infty}$ and $U$ in 
\cite{Maldacena:2009mw} as follows: 
\[
h_1=2\gamma^2 \widetilde{N}_c,\;\;\; c=\frac{\widetilde{N}_c}{6}e^{-\frac23t_{\infty}},\;\;\;
U  =\frac{2\widetilde{N}_c}{c}.
\label{identificationofparameters}
\] 
Moreover, $\tilde N_c$ here is $\tilde M$ in \cite{Maldacena:2009mw}. 
The complete map to the variables used in 
\cite{Maldacena:2009mw} includes:  $t_{MM}=2\rho$, $\tau_{MM}=t$, 
$c_{MM}=P/{\tilde N_c}$, and $f_{MM}=4{\cal P}/{\tilde N_c}$. 
See also Appendices \ref{more-on-solutions} and \ref{mmvar}.}
\be\label{MM-asymptotics}
P=\left\{\begin{matrix}
          h_1 \r+ \frac{4 h_1}{15}\left(1-\frac{4 \widetilde{N}_c^2}{h_1^2}\right)\r^3
+\frac{16 h_1}{525}\left(1-\frac{4\widetilde{N}_c^2}{3h_1^2}-
\frac{32\widetilde{N}_c^4}{3h_1^4}\right)\r^5+\co(\r^7), & \r\to 0, \\ & \\
c e^{4\rho/3}+\frac{4\widetilde{N}_c^2}{c}\left( \r^2-\r+\frac{13}{16}\right)e^{-4\rho/3}+\co(\r e^{-8\rho/3}), & \r\to\infty.
         \end{matrix}\right.
\ee
In the full solution the two constants are related in a non-trivial way 
\cite{Maldacena:2009mw}, given in (\ref{UV-IR-relation}) of  Appendix \ref{more-on-solutions}. 
What is important for the present discussion is that $h_1(c)$ as
a function of $c$ takes values in $[2,+\infty)$, while $c\in [0,+\infty)$, with $h_1(0)=2$. 
One can also construct the solution
in an expansion for large $c$, as is discussed in detail in Appendix \ref{more-on-solutions}. 
One then finds,  via (\ref{BPS-sols}), that the dilaton takes the form
\be
e^{2\F}=e^{2\F_\infty}\left(1-\frac{1}{c^2}h_{KS}(\r)+\co(1/c^4)\right),
\ee
where 
\be
e^{2\F_\infty}\equiv\sqrt{\frac32}\frac{e^{2\F_o}}{c^{3/2}},
\label{dildil}
\ee
and
\be
h_{KS}=2^{1/3}\widetilde{N}_c^2\int_{\r}^\infty\frac{d\r'}{\sinh^2(2\r')}\left(2\r'\coth(2\r')-1\right)\left(\sinh(4\r')-4\r'\right)^{1/3},
\ee
is the Klebanov-Strassler warp factor (cf. eq.~(90) in \cite{Klebanov:2000hb}). 

Let us recall  some limits of this two-parameter family of solutions discussed in \cite{Maldacena:2009mw}:

\begin{itemize}

\item{\em $\b\to 0$}

This is the original  background before adding the D3 brane charge \cite{Casero:2006pt}, namely it describes \emph{wrapped D5 branes}.
The interpretation of the parameter $c$ was discussed in  
\cite{HoyosBadajoz:2008fw},\cite{Maldacena:2009mw}.  
Taking $c\to 0$ corresponds to going to the near-brane limit, which is the solution discussed 
in \cite{Maldacena:2000yy}. In this decoupling limit the theory on the fivebranes was argued to flow to
pure ${\cal N}=1$ in the IR \cite{Maldacena:2000yy}. In the opposite $c \to \infty $ limit,
 the metric approaches the deformed conifold with three-form  flux.

\item{\em  $\b\to \infty$}

This limit ensures that the constant term in the warp factor in (\ref{newwarp}) is removed and 
the leading term in the UV is dominated by $h_{KS}$. The expansion in large $c$ does not
terminate and $c$ remains as the only non-trivial parameter of the solution in addition to 
$\F_\infty$. This solution describes the  \emph{baryonic branch} of the Klebanov-Strassler theory
and the parameter $c$ is related to the baryonic branch VEV as $U\propto c^{-1}$ \cite{Butti:2004pk}.

\item {\em $\b \to \infty $ and $c\to \infty$}

In this case the warp factor $\hat{h}$ in (\ref{rotated-background}) is 
replaced by $h_{KS}$ and hence the background is the \emph{Klebanov-Strassler} background  \cite{Klebanov:2000hb}. In particular, the 
unwarped internal metric is the deformed conifold metric. 
The only free parameter  is the asymptotic value of the dilaton $\F_\infty$. 
The deformation parameter $\epsilon$ of the deformed conifold  may be 
reabsorbed by a rescaling of the metric.

\item{\em $\b \to \infty $ and $c\to 0$}

This is the limit of large VEVs on the baryonic branch solution.
In  \cite{Maldacena:2009mw} it was shown that there is
a large region where the dilaton is approximately constant and the 
solution is well approximated by the solution  of \cite{Pando Zayas:2000sq}.
It was then argued that in this limit one approaches the wrapped fivebrane theory, 
but with a $B$ field on the two-sphere, induced by the ``rotation'' procedure. Therefore in this case 
the theory is well described by fivebranes wrapped on a \emph{fuzzy two-sphere}.

\end{itemize}

\subsubsection{The new flavoured solution}
\label{newflavor}

\begin{figure}[ht]
\begin{minipage}[l]{0.5\linewidth}
\centering
\includegraphics[scale=.8]{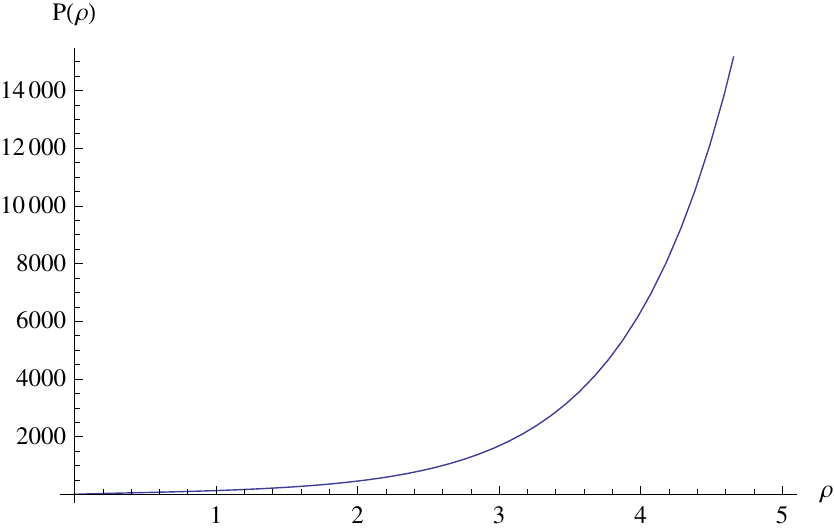}
\end{minipage}
\hspace{0.5cm}
\begin{minipage}[l]{0.5\linewidth}
\centering
\includegraphics[scale=.8]{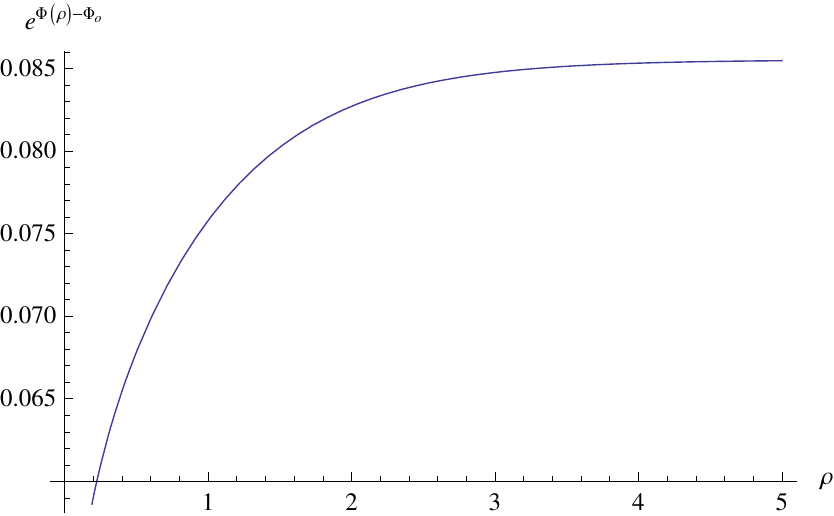}
\end{minipage}
\caption{Plot of the function $P(\r)$  and the dilaton for the numerical solution interpolating between 
the two asymptotic behaviours in (\ref{new-solution-asymptotics}). The plots correspond to the values
$\widetilde{N}_c=10$, $\widetilde{N}_f=20$, $c=30$, and $h_1=100$.   
}
\label{plots-flavored}
\end{figure}

\begin{figure}[ht]
\begin{minipage}[l]{0.5\linewidth}
\centering
\includegraphics[scale=.8]{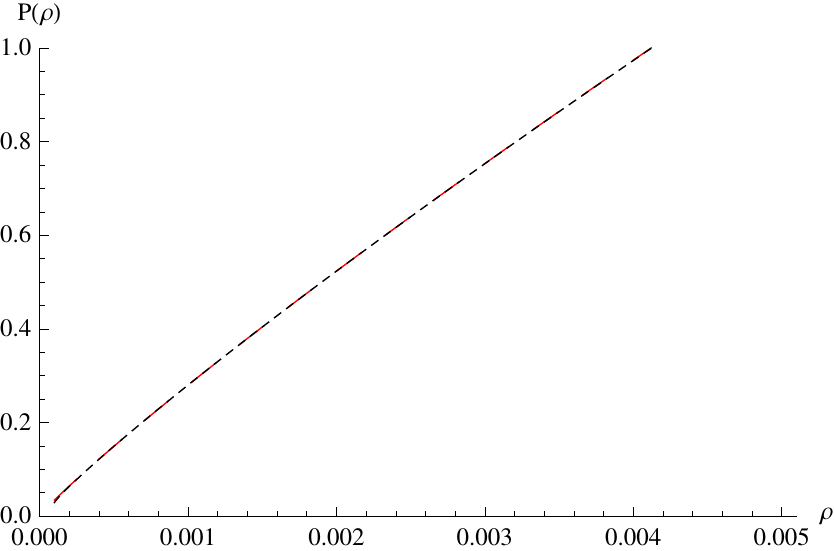}
\end{minipage}
\hspace{0.5cm}
\begin{minipage}[l]{0.5\linewidth}
\centering
\includegraphics[scale=.8]{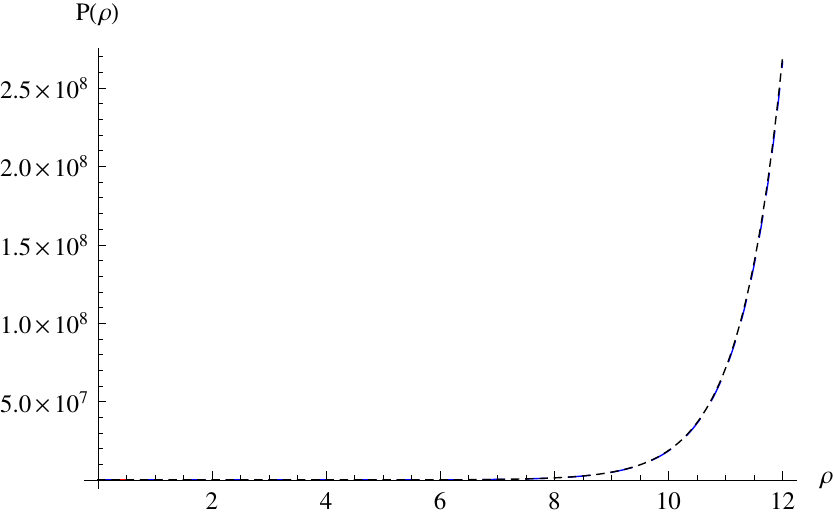}
\end{minipage}
\caption{In these plots we plot the same numerical solution as in Fig. \ref{plots-flavored}, but we zoom
in on the IR region (left) and on the UV region (right) and we compare the numerical solution (black) with the corresponding asymptotic
solutions given in (\ref{new-solution-asymptotics}). These are plotted in red (IR solution) and in blue (UV solution). }
\label{flavored-asymptotics-vs-numerics}
\end{figure}

Let us now present the new flavoured resolved deformed conifold 
solution. Note that the analysis so far is general
enough to allow for smeared D5 sources and therefore
 we only need to find a new solution of the ``master equation''
(\ref{master}) with $\widetilde{N}_f\neq 0$, subject to the condition that it reduces to the solution of
\cite{Butti:2004pk} in the limit $\widetilde{N}_f\to 0$. 
We have found this new solution numerically, but again one can systematically determine the IR and UV
asymptotics, discussed in detail in Appendix \ref{more-on-solutions}: 
\be\label{new-solution-asymptotics}
P=\left\{\begin{matrix}
          h_1\r+\frac{4\widetilde{N}_f}{3}\left(-\r\log\r-\frac{1}{12}\r\log(-\log\r)+\co\left(\frac{\r\log (-\log \r)}{\log \r}\right)\right)
+\co(\r^3\log \r), & \r\to 0, \\ & \\
c e^{4\r/3}+\frac{9\widetilde{N}_f}{8}+\frac{1}{c}\left((2\widetilde{N}_c-\widetilde{N}_f)^2
\left(\r^2-\r+\frac{13}{16}\right)-\frac{81\widetilde{N}_f^2}{64}\right)e^{-4\r/3} +\co(\r e^{-8\rho/3}), & \r\to\infty.
         \end{matrix}\right.
\ee 
These solutions are different from those discussed in 
\cite{HoyosBadajoz:2008fw} and in particular they 
reduce to their unflavoured counterparts in (\ref{MM-asymptotics}) in the 
limit $\widetilde{N}_f\to 0$. 
However, contrary to the unflavoured solution, the flavoured solution is singular in the IR.  
Moreover, note that the flavours, \emph{i.e.} terms proportional to $\tilde{N}_f$, 
dominate the IR, as well as the UV when $\b\to\infty$ (see (\ref{baryonic-branch-warp}) below). 
Via (\ref{BPS-sols}) we see that this asymptotics imply
 that the dilaton goes to $-\infty$ in the IR and not to a constant as in the unflavoured case. In particular, 
for small values of the radial  coordinate we have\footnote{To recover the limit $\tilde N_f=0$ at fixed $c$ the expansions
 (\ref{thedilat}) and (\ref{irwarpo}) are not useful. This limit is completely smooth as can be seen from the expansions of $P$ in (\ref{new-solution-asymptotics}) and the various plots of the numerical solutions.}
\be
e^{2\F}=\frac{3e^{2\F_\infty}c^{3/2}}{\tilde N_f^{3/2}(-\log\r)^{3/2}}\left(1-\frac{\log(-\log\r)}{8\log\r}
+\co\left(\frac{1}{\log\r}\right)\right).
\label{thedilat}
\ee
Thus the warp factor has the following expansion in the IR
\bea
\hat{h}=\frac13e^{-2\F_\infty}\left(\frac{2\tilde N_f}{c}\right)^{3/2}(-\log\r)^{3/2}
\left(1+\frac{\log(-\log\r)}{8\log\r}
+\co\left(\frac{1}{\log\r}\right)\right).
\label{irwarpo}
\eea
The numerical solution interpolating between the asymptotic behaviours in (\ref{new-solution-asymptotics}) is
plotted in Fig. \ref{plots-flavored}. In Fig. \ref{flavored-asymptotics-vs-numerics} the numerical
solution is explicitly compared with the asymptotic solutions (\ref{new-solution-asymptotics}) by zooming 
in the IR and UV regions. 

Let us now reconsider the various limits discussed in the previous section for the unflavoured solution: 

\begin{itemize}

\item{\em $\b\to 0$}

Again, this is the original D5 brane background before adding the D3 branes.   
Further taking $c\to 0$ is the near-brane (decoupling) limit.
In this case, the resulting solution interpolates between  the IR asymptotics given 
in (\ref{new-solution-asymptotics}) and the 
linear dilaton asymptotics $P\sim 
|2\widetilde{N}_c-\widetilde{N}_f|\r+P_o$ (with $P_o=\frac{\widetilde{N}_f}{2}$).
The solution is plotted in Figs. \ref{plots-flavored-linear-dilaton} and 
\ref{flavored-asymptotics-vs-numerics-linear-dilaton}
and it is the flavoured generalisation of the wrapped D5 solution of \cite{Maldacena:2000yy}.

\item{\em $\b\to \infty$}

Although in this limit we remove the constant term from the warp factor in the UV, now the leading form of 
the warp factor is not dominated by $h_{KS}$ any more, but by the term introduced by the sources. 
In particular, using the expression (\ref{dilaton-expansion}) for the 
dilaton we have
\be\label{baryonic-branch-warp}
\hat{h}=e^{-2\F_\infty}\left(\frac{2^{2/3} 
\widetilde{N}_f}{c}\int_\r^\infty 
d\r'\left(\sinh(4\r')-4\r'\right)^{-1/3}+\co(1/c^2)\right).
\ee
It follows that the UV asymptotic behaviour of the flavoured  
solution is \emph{different} from the Klebanov-Strassler asymptotics.  As we will see later, 
this leads to a different field theory picture.

\item {\em $c\to \infty$ and $\b\to \infty$ }

This limit cannot be taken naively in the flavoured case, due to the fact 
that the flavours dominate the UV after the leading constant term in the 
warp factor  is removed. The reason why we cannot go to the Klebanov-Strassler limit, 
while keeping the flavour D5 branes, is that this is not a supersymmetric configuration \cite{Martucci:2006ij}. 
Therefore we will consider the limit of $c\to \infty$ and $\widetilde{N}_f\to 0$ at fixed $c\widetilde{N}_f$.
This limit will be the subject of the next subsection. 
In section \ref{comme}, we will argue that the field theory interpretation of 
this solution is a modification (by Higgsing)  of the Klebanov-Strassler cascade.

\item{\em $\b \to \infty $ and $c\to 0$}

One can perform an analysis similar to that in 
\cite{Maldacena:2009mw} for the  solution in this range of parameters and show that 
there is again a large region where the solution is well approximated by a resolved conifold 
metric, with addition of fluxes \emph{and} sources. This suggests that there should exist an exact Type IIB
solution analogous to that in \cite{Pando Zayas:2000sq}, modified by the presence of sources. 
It would be interesting to find this solution.

\begin{figure}[ht]
\begin{minipage}[l]{0.5\linewidth}
\centering
\includegraphics[scale=.8]{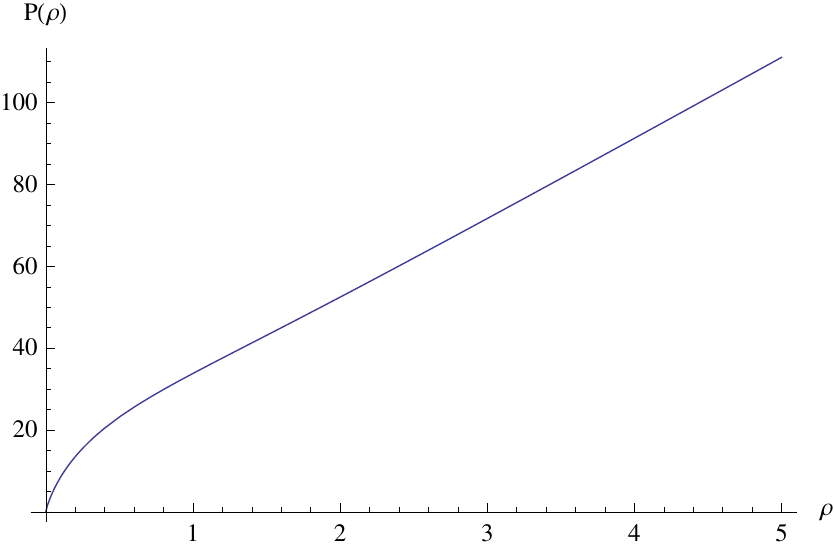}
\end{minipage}
\hspace{0.5cm}
\begin{minipage}[l]{0.5\linewidth}
\centering
\includegraphics[scale=.8]{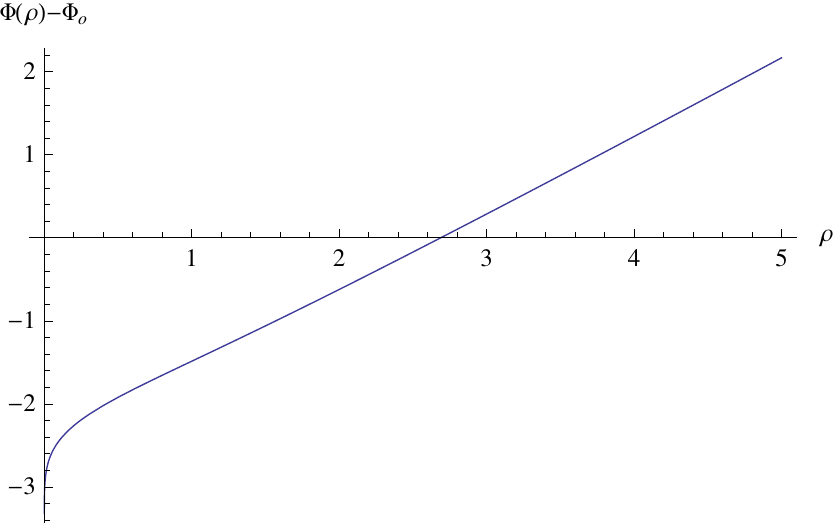}
\end{minipage}
\caption{Plot of the function $P(\r)$  and the dilaton for the numerical solution interpolating between 
the IR asymptotic behaviour in (\ref{new-solution-asymptotics}) and 
the linear UV asymptotics $P\sim |2\widetilde{N}_c-\widetilde{N}_f|\r$. 
The plots correspond to the values $\widetilde{N}_c=20$, $\widetilde{N}_f=20$, $P_o=10$ 
and $h_1=25.93$. Contrary to the $\widetilde{N}_f=0$ case, we do 
not have an analytic expression for the value of $h_1$ leading to linear 
dilaton asymptotics in the UV.  }
\label{plots-flavored-linear-dilaton}
\end{figure}
\begin{figure}[ht]
\begin{minipage}[l]{0.5\linewidth}
\centering
\includegraphics[scale=.8]{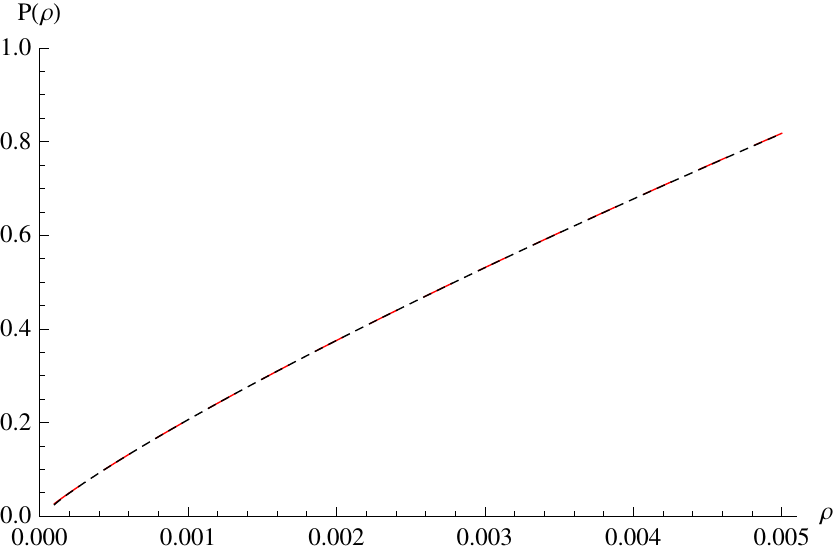}
\end{minipage}
\hspace{0.5cm}
\begin{minipage}[l]{0.5\linewidth}
\centering
\includegraphics[scale=.8]{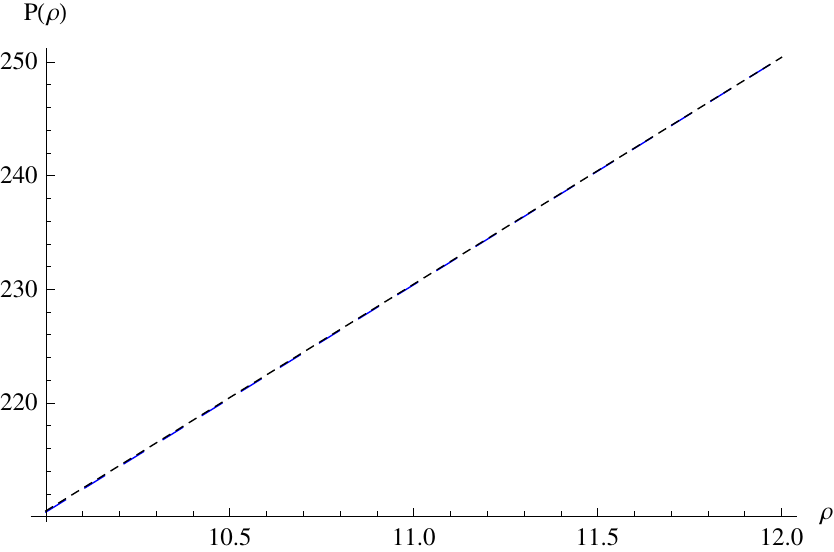}
\end{minipage}
\caption{Here we plot the same numerical solution as in Fig. 
\ref{plots-flavored-linear-dilaton}, but we zoom
in on the IR region (left) and on the UV region and we compare the numerical solution (black) with the IR asymptotic
solution given in (\ref{new-solution-asymptotics}) and the UV asymptotic solution $P\sim |2\widetilde{N}_c-\widetilde{N}_f|\r$.  
These are plotted in red (IR solution) and in blue (UV solution). }
\label{flavored-asymptotics-vs-numerics-linear-dilaton}
\end{figure}

\end{itemize}

\begin{flushleft}
{\em Summary}
\end{flushleft}

\begin{figure}[h]
\begin{minipage}[l]{0.5\linewidth}
\centering
\includegraphics[scale=.8]{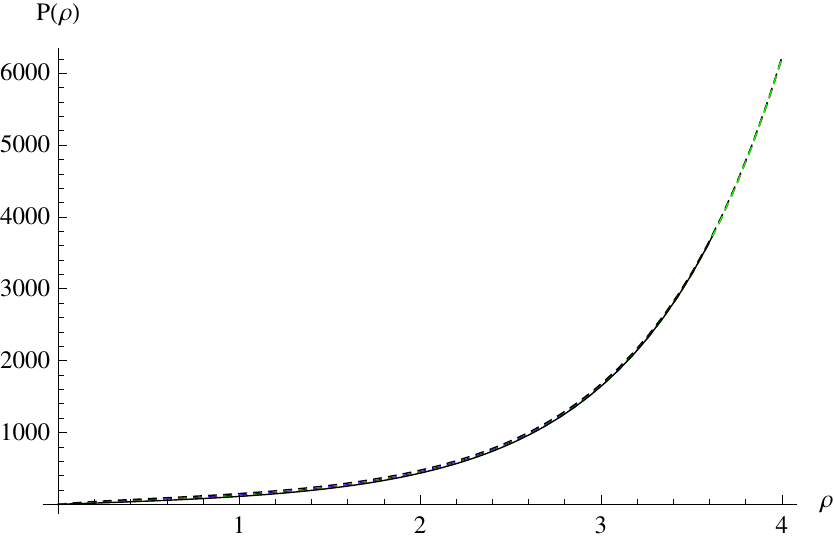}
\end{minipage}
\hspace{0.5cm}
\begin{minipage}[l]{0.5\linewidth}
\centering
\includegraphics[scale=.8]{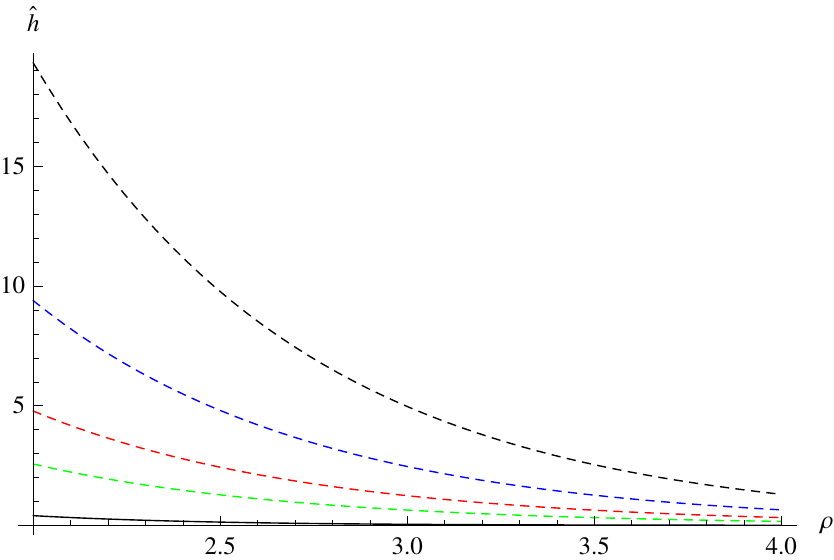}
\end{minipage}
\caption{On the left: plots of  $P(\rho)$ for fixed values $c=30$, $\widetilde{N}_c=10$ and different values of $\widetilde{N}_f$ (and $h_1$).
The continuous curve is $\widetilde{N}_f=0$. Superimposed on this are the curves for 
the following values:
$\widetilde{N}_f=5$ (dotted green), $\widetilde{N}_f=10$ (dotted red), $\widetilde{N}_f=20$,  (dotted blue), $\widetilde{N}_f=40$ (dotted black). 
On the right: different plots of $\hat{h}(\rho)$  for the same values of $\widetilde{N}_f$.}
\label{summaryplots}
\end{figure}

Let us  summarise the effects of the addition of the flavour D5 branes to the unflavoured solution. 
The ansatz for the metric and fluxes is essentially unchanged and may be parameterised completely in terms
of the function $P$ - see Appendix \ref{mmvar} for a concise presentation of the ansatz. 
The function  $P$ for various values of $\widetilde{N}_f$ is plotted on the left in  Figure \ref{summaryplots}.
Notice that the leading UV behaviour of $P$ is not affected by the 
flavours. However, the flavoured solution is actually  singular at $\rho=0$. 
 After introducing the D3 branes, the six-dimensional  metric $\dd s^2_6$ is unchanged, 
and   is warped by  the warp factor (\ref{newwarp}). This picks up the  sub-leading behaviour of the dilaton in the UV and therefore 
it is sensitive to the $\widetilde{N}_f$ flavours - see the plots of $\hat{h}$ on the right of Figure  \ref{summaryplots}. 
The divergence of  $\hat {h}$ at $\rho=0$ is due to the singularity in the IR.
The fall-off  at infinity is noticeably slower with respect to the unflavoured case, and we expect that this 
will persist after resolving the IR singulatity. To understand the physical 
origin of the UV  behaviour, we will next discuss the solution in a limit in which the six-dimensional
(unwarped) metric becomes an ordinary  deformed conifold.

\subsection{Adding smeared D3 branes to the Klebanov-Strassler theory}
\label{flavKS}

We now discuss a solution obtained in the limit $c\to\infty$ (with $\tanh \b =1$) of the 
flavoured solution.  This limit can be obtained by inserting the 
warp factor (\ref{baryonic-branch-warp}) in (\ref{newwarp}) and sending $c\to\infty$. 
However, the fact that $\hat{h}\sim 1/c$ and
not $\hat{h}\sim 1/c^2$, as is the case for the unflavoured solution, does not allow to take 
this limit directly.  To obtain a well-defined limit we set
$c\widetilde{N}_f=\nu$, and keep  $\nu$  fixed in the limit $c\to\infty$. 
Using (\ref{dilaton-expansion}), we obtain 
\be
\hat{h}=\frac{1}{c^2}e^{-2\F_\infty}\left(2^{2/3}\nu\int_\r^\infty 
d\r'\left(\sinh(4\r')-4\r'\right)^{-1/3} +h_{KS}\right)+\co(1/c^3).
\ee
Inserting this in (\ref{rotated-solution}) and sending $c\to\infty$ we 
obtain an exact solution (notice that the expressions for $B$ and $F_3$ below do not scale with the parameter $c$) 
\bea
\label{exact-flavored-solution}
&&\dd s_{str}^2=e^{\Phi_\infty} \left[ h_{\nu}^{-1/2}\dd x_{1,3}^2+h_{\nu}^{1/2}\dd s_6^2(\widetilde{N}_c,0)\right]\NO\\
&&F_3=\ast_6 \dd J^{(0)}(\widetilde{N}_c,0),\NO\\
&&B=e^{\F_\infty}J^{(0)}(\widetilde{N}_c,0),\NO\\
&&F_5=-e^{-\F_\infty}(1+\ast_{10})\dd h_{\nu}^{-1}\wedge \vol_{(4)},
\eea
where  $\dd s_6^2(\widetilde{N}_c,0)$ is the \emph{deformed conifold}
metric and we defined
\be
h_{\nu} = 2^{2/3} \nu \int_\r^\infty 
d\r'\left(\sinh(4\r')-4\r'\right)^{-1/3} +  h_{KS}.
\label{newWF}
\ee 
To understand the significance of this solution, notice that $\nu \neq 0$ leads to 
\be
\dd F_5-H_3\wedge F_3=B\wedge \X_{(4)}\neq0 ,
\label{spqr}
\ee
where the  term on the right-hand side of this equation
 may be interpreted as the contribution from  D3 brane sources smeared on the transverse directions.
 Indeed, the world-volume action for these sources arises in this  limit, as discussed in subsection \ref{limitsection}.

Going back to the solution \emph{before} taking the $c\to \infty$ limit, 
we can think of the  term $B \wedge \Xi_{(4)}$ in the second equation 
in (\ref{f5bianchi}) as a  D3 brane charge density induced on the source D5 branes by the $B$ 
field on their world-volume.  Then we can compute the density of D3 branes by integrating the $B$ field pulled back to the 
world-volume cylinder wrapped by the D5 branes, where we put a cut-off at 
some radial distance.  The result
is also valid in the $c\to \infty$ limit.
Namely, we may define a running 
number\footnote{In the following formulas we will ignore numerical factors.}  of source D3 branes as 
\bea
n_f  = \frac{\widetilde{N}_f}{(2\pi)^2}\int_{\mathrm{cylinder}} \!\!\!\! B  ~\propto~ \widetilde{N}_f e^{-\Phi_\infty}\int^\rho e^{2\Phi+2k} d\rho',
\eea
where the factor of $\widetilde{N}_f$ comes from the overall factor in front of the action for the flavour D5 branes.
Expanding this in the UV we get ($g_s=e^{\Phi_\infty}$)
\bea
n_f \sim g_s \nu e^{4\rho/3} ~\qquad \mathrm{for}~~~~\rho \to \infty.
\eea 
The interpretation of this quantity becomes clear if we look at the asymptotic form of the warp factor
in the standard radial coordinate  $r\sim e^{2\rho/3} $.  The leading term of the warp factor  
in the UV goes like 
\bea
  h_\nu    \sim  \frac{\nu r^2 + \widetilde{N}_c^2 \log r } {r^4} \qquad \mathrm{for}~~~~r \to \infty.
\eea
Expressing this in terms of the  running number of source 
D3 branes $n_f$,  and running number of bulk 
D3 branes  \cite{Klebanov:2000hb},
\bea
n \sim k \widetilde{N}_c \sim  g_s \widetilde{N}_c^2 \log r ~ ~\qquad \mathrm{for}~~~~ r \to \infty,
\eea
this takes the form
\bea
g_s h_\nu \sim  \frac{n_f + n}{r^4} \qquad \mathrm{for}~~~~r \to \infty.
\label{uvwarp}
\eea
This shows that  there are precisely $n_f+n$ D3 branes in the background and reduces
to the  Klebanov-Strassler expression for $\nu=0$. Notice that  
the running of the source and bulk D3 branes is quite different, and in 
particular the former dominates  the UV. We will discuss later the 
implications of this behaviour. 
In fact, this is the leading asymptotic behaviour of the warp factor (\ref{baryonic-branch-warp})
in the general case.

The limiting solution is again singular near $\rho=0$. However, this singularity comes entirely from the warp factor 
(\ref{newWF}), while the metric $\dd s_6^2(\widetilde{N}_c,0)$ is the smooth deformed conifold metric.
This singularity is due to the fact that the D3 sources are distributed uniformly along the radial direction down to $\rho=0$.


\section{Comments on the field theory}
\label{comme}

In this section we will discuss possible field theory interpretations of the solutions we presented.
As we will see, the solution of section \ref{flavKS} will be related to the mesonic branch 
of the Klebanov-Strassler theory, whereas a detailed field  theory picture for the solutions 
with non-zero $\widetilde{N}_f$ and $c^{-1}$  is more difficult to obtain. 
We find instructive to first recall the relation 
between the gravity solutions and the field theory picture in the 
unflavoured theories, as discussed in \cite{Maldacena:2009mw}. 
In that case the fivebrane solution contains one non-trivial parameter $U\propto c^{-1}$.  
Taking this to infinity corresponds to the near-brane limit,  where a decoupled four dimensional field theory
description exists. For any finite, non-zero value of $U$ 
we can ``rotate'' the solution and add D3 branes and a new parameter $\beta$. In the new solution, 
the decoupling limit is obtained by taking $\beta \to \infty$. Thus $U$ survives as a 
field theory parameter, and in particular it is proportional 
to the VEV of the ${\cal U}$ operator, partner of the conserved baryonic current.

As already noted, the 
flavoured fivebrane solution has the same UV asymptotics as the unflavoured one\footnote{Only the sub-leading behaviour is affected.} 
and is still charaterised by 
a parameter, denoted by $c$. The decoupling (near-brane) limit is obtained 
by sending  $c\to 0$, where  the asymptotic dilaton is linear, instead of constant.  
After the transformation that adds
the D3 branes (for generic values of $c$), the warp factor 
will asymptote to a constant, unless again we take 
$\beta \to \infty$. We would like then to regard this as a limit in which a 
four dimensional  field theory interpretation should be possible. 
 To be more precise, let us write the leading expansion of the warp factor
in the UV, after having dropped the constant term. Using the expansions in 
Appendix \ref{more-on-solutions}, we get
\bea
\hat{h}=\frac{e^{-2\F_\infty}}{c^2}\left(\frac{3c\widetilde{N}_f}{2}e^{-4\r/3}
+\frac{3}{32}\left[(2\tilde{N}_c-\widetilde{N}_f)^2(8\r-1)-297\widetilde{N}_f^2\right]e^{-8\r/3}+\co(e^{-4\r})\right)
\label{pity}
\eea
and changing to the radial variable $r\sim e^{2\rho/3}$ we have
\bea
 \hat  h  =  \frac{3e^{-2\F_\infty}}{2c^2}  \left(\frac{\nu r^2 +  3(\widetilde{N}_c-\widetilde{N}_f/2)^2 \log r } {r^4}\right)  +\co(1/r^4) .
\label{truewarp}
\eea
Thus in terms of the effective number of running D3 branes
\bea
  n_f \sim g_s \nu r^2 ~~~~~~~~\mathrm{and}~~~~~~~~ n \sim g_s (\widetilde{N}_c-\widetilde{N}_f/2)^2 \log r ~~~~~~~~~\mathrm{for}~~~~ r \to \infty,
\eea
we see that again\footnote{The factors of the dilaton combine with the factor in front of 
$e^{2\Phi}\dd s_6^2(\widetilde{N}_c,\widetilde{N}_f)$ in (\ref{rotated-solution}). 
The factor of $c^{-2}$ is canceled by an overall factor of $c$ 
in the asympotic  expression of 
$\dd s_6^2(\widetilde{N}_c,\widetilde{N}_f)$, so that the final metric does not depend on $c$ at infinity.} 
\bea
g_s \hat h \sim  \frac{n_f + n}{r^4} \qquad ~~~\mathrm{for}~~~~~r \to \infty.
\label{genuvwarp}
\eea
Comparing with the limiting solution discussed in section \ref{flavKS}, we 
see that 
the term proportional to $n_f$ is due to the smeared source D3 branes, while the subleading term is 
due to the running bulk  D3 branes. In particular, the former leading term does  not give 
the usual asymptotic 
AdS  with logarithmically running radius. 
This might suggest that a  4d field theory interpretation may not be available 
for these solutions. However, 
we can define the theory at some energy scale, \emph{i.e.} 
at some radial distance $r_*$. 
Once the field theory is defined, we may then follow its evolution in the far UV, as for the 
ordinary Klebanov-Strassler  theory. Notice the present discussion is valid both for 
vanishing $\widetilde{N}_f$ or otherwise.

The difference in asymptotics  with respect to the usual logarithmic deviation from 
AdS$_5\times T^{1,1}$  has an important implication in the dual field theory. In particular, 
we have that the would-be Goldstone boson  fluctuation \cite{Gubser:2004qj} 
around the solution of section \ref{flavKS}  is \emph{not normalisable}. 
This is easy to check by noticing that the ansatz in \cite{Gubser:2004qj}
is still valid in our case, and that the first integral in (3.27) 
of  \cite{Gubser:2004qj} is divergent  with our asymptotic behaviour of the warp factor.  
Consequently, in our case we do not find a 
Goldstone boson associated with spontaneous breaking of the global 
baryonic symmetry 
$U(1)_B$. Therefore $U(1)_B$ is not broken\footnote{The absence of a Goldstone boson might be due 
to the fact that $U(1)_B$ is gauged. However this seems
unlikely for our solutions. We thank Ofer Aharony for pointing this out to us.}, 
suggesting that our solutions always describe non-baryonic branches. 
The mode we are discussing corresponds to (infinitesimally) turning on the resolution parameter in the geometry 
(\emph{i.e.} in the unwarped metric), which in our solution corresponds to turning on $c^{-1}$.  
The fact that $U(1)_B$ is not broken strongly suggests that  
the operator  ${\cal U}$, defined as the partner of the conserved $U(1)_B$ current, 
\emph{does not} have a VEV, hence we cannot associate the parameter 
$c^{-1}$ in the gravity to this VEV.  In fact, in the present situation
this parameter  may be  interpreted simply as due to the back-reaction of the flavour D5 branes. 
In formulas, we have that $c^{-1} = \widetilde{N}_f/\nu$, thus we see that 
introducing $\widetilde{N}_f$ D5 branes wrapped on a cylinder transverse to the $S^2$
of the resolved conifold  turns on a finite size (as measured from infinity - see \cite{Maldacena:2009mw}) 
for this two-sphere. More precisely, the parameter $c^{-1}$ is proportional to the ratio of D5 to D3 sources. 
Notice that the D5 brane sources break the $\Z_2$ symmetry of the Klebanov-Strassler  solution. We will elaborate on this point in 
section \ref{thez2}.

Based on the above comments, we now turn to discuss more precise field theory duals to the backgrounds
 we presented.  We will start from the solution of section \ref{flavKS} and then we will move on to the general 
solution. In this case, we will examine two distinct possibilities,  pointing out pros and cons of both scenarios.

\subsection{Higgsing in the Klebanov-Strassler theory}
\label{deformedzzz}

The solution\footnote{A closely related supergravity solution
was discussed briefly in  \cite{Klebanov:2000hb}, although notice that 
the warp factor in our equation (\ref{newWF}) differs from 
the corresponding one in equation (101) of \cite{Klebanov:2000hb}. 
Indeed the latter is a zero-mode of the 
source-less Laplace equation on the deformed conifold.}
discussed  in section \ref{flavKS} 
consists of an ordinary deformed conifold, 
with a warp factor comprising the standard 
source-less term $h_{KS}$, whose origin is the running number $n$ of bulk D3 branes, 
plus a contribution arising from  $n_f$ \emph{smeared} source D3 branes. 
We therefore propose that the field theory dual to this 
solution is the Klebanov-Strassler theory with gauge group\footnote{From now on we drop the tilde 
 from $\widetilde N_c$, $\widetilde N_f$ and denote them simply as $N_c$, $N_f$.}
\bea
 SU(N_c+n+n_f)\times SU(n+n_f)
\label{HKS}
\eea
on the \emph{mesonic branch}. 
To understand this more precisely, let us first recall the structure
 of the moduli space of this theory, as discussed in detail in 
\cite{Dymarsky:2005xt}. 
In the notation of  this reference 
 the generic Klebanov-Strassler quiver has gauge group
\bea
SU(M(k+1)+\tilde p)\times SU(kM+\tilde p).
\label{DKSquiver}
\eea
The authors of \cite{Dymarsky:2005xt} showed that the
 (quantum) moduli space of this theory consists 
of different branches, and may be written as a direct sum
\bea
\oplus_{r=1}^M \left[\Sym^{p} {\cal C}_{r,0} \oplus \Sym^{p-M} {\cal C}_{r,1} \oplus \Sym^{p-2M} {\cal C}_{r,2} \oplus  \cdots \oplus \Sym^{\tilde p}~{\cal C}_{r,k} \right]
 \label{DKSbranches}
\eea
where $p=kM+\tilde p$.
Each of the ${\cal C}_{r,l}$ denotes a deformed conifold, with deformation parameter depending on 
two integers $r,l$. The sum over $r=1,\dots , M$ is related to the $M$ vacua arising in each of the terms, while the  
integer $l$ labels branches of the moduli space arising at different steps of the cascade of Seiberg dualities. 
The power of the symmetric product is interpreted as the number 
of  \emph{probe}  D3 branes moving on the corresponding deformed conifold 
geometry. The last term in the sum is the branch of lowest dimension, that is real dimension $6\tilde p$.
However, if $\tilde p=0$, the branch of lowest dimension is replaced by the \emph{baryonic branch}, 
which is a copy of $\mathbb{C}$ \cite{Dymarsky:2005xt}.

Let us now consider a supergravity solution which incorporates the back-reaction 
 of  $n_f$ point-like D3 branes at distinct points on a warped deformed conifold geometry. 
This is different from the solution of section \ref{flavKS}.
The back-reaction of the source D3 branes is included 
by replacing the warp factor  $h_{KS}\to h_{KS} +\sum_{j=1}^{n_f} h_j$, 
where each $h_j$ is a solution 
to the Laplace equation on the deformed conifold,  with a delta-function source. 
If the branes are placed  at some point on the $S^3$ at $\rho=0$,
we can  interpret this solution\footnote{A solution corresponding to coincident 
D3 branes localised  at a particular point on the deformed conifold was discussed in 
\cite{Krishnan:2008gx}.}  in terms of a cascade of Seiberg 
dualities of the theory (\ref{DKSquiver}) 
down to the last step, where it becomes $SU(M+\tilde p)\times SU(\tilde p)$ and 
then it goes over to the smallest branch of (\ref{DKSbranches}),  with $\tilde p = n_f$. 
However, if the D3 brane sources are placed at some finite radial distance, 
the natural holographic interpretation 
of the background is that the theory first undergoes a cascade of Seiberg dualities, and then 
it is Higgsed  at an energy scale given by this distance. 
Then we are in some intermediate branch in (\ref{DKSbranches}).

Let us now return to the solution of section \ref{flavKS}, 
corresponding to smearing uniformly the $n_f$ source D3 branes on the transverse geometry.
Every time a D3 brane is crossed, the ranks of both gauge groups decrease by one unit. 
We  may then interpret the gravity solution as a continuous process of  
\emph{Higgsing},  occurring at all energy scales from the  UV to the IR. 
Part of the change in the flux, in the geometry,  and in the ranks in the 
field theory, is then due to this Higgsing, and is reflected in the 
$\nu$ term in the warp factor. However, we still have the logarithmic part of the running, exactly as in Klebanov-Strassler, 
implying that the theory is also undergoing a cascade of Seiberg dualities. 
After resolving the singularity, it may be possible to stop the Higgsing at 
some finite distance, and proceed with the cascade.  However,   in our singular solution 
the Higgsing behaviour is dominating down to the IR, and therefore it seems that 
 the Higgsed theory in the IR should be $SU(N_c)$, where all of the D3 branes have disappeared. 
This possibility was entertained in \cite{Aharony:2000pp}, who concluded that ultimately 
for the solution of \cite{Klebanov:2000hb} the cascade interpretation is the correct one.  
On the other hand, in the presence of smeared sources like in our solutions, 
the Higgsing interpretation appears inevitable  \cite{Aharony:2000pp}.

Standard supergravity computations, as for example  
the beta functions  of the two gauge couplings,  defined  exactly as in \cite{Klebanov:2000hb}, 
can be matched to the field theory computations, and are 
compatible  with our interpretation. In summary, we have proposed that the gravity solution 
of  section \ref{flavKS} corresponds to the mesonic branch of 
the Klebanov-Strassler theory with gauge groups given in 
(\ref{HKS}),  simultaneously undergoing a cascade of Seiberg dualities and  Higgsing.

\subsection{Higgsing in the Klebanov-Strassler theory with flavours}
\label{speculations}

Let us now address switching on a non-zero value for $N_f$.
One way to think about the final background is as 
arising from the back-reaction of $N_f$ D5 branes\footnote{Reference \cite{Dymarsky:2009fj} studied 
placing flavour D7 branes in the resolved deformed conifold geometry.}
placed in the resolved deformed conifold geometry of \cite{Butti:2004pk}. 
However, $\kappa$-symmetry requires to turn on a world-volume flux on the D5 branes \cite{Martucci:2006ij}. 
Thus we have necessarily source D3  branes induced on the flavour D5 branes.
According to this point of view, the most direct interpretation of the 
solution 
with $N_f$ and $c^{-1}$ non-zero seems to be 
in terms of flavour D5 branes on the \emph{baryonic branch}  of the $SU((k+1)N_c)\times SU(kN_c)$ 
Klebanov-Strassler theory. However, we cannot add $N_f$ flavour D5 branes 
 without also adding extra (infinite) $n_f$ D3 branes.
On the other hand, from the point of view of the picture presented 
in section \ref{deformedzzz},
turning on $N_f$ and $c^{-1}$ looks like a deformation of the \emph{mesonic branch} of 
the $SU(N_c+n+n_f)\times SU(n+n_f)$ Klebanov-Strassler theory in  (\ref{HKS}).   
Below we will analyse two different scenarios 
for the dual field theory of our solution, based on these two different viewpoints, respectively.

\subsubsection{Explicit flavour symmetry}
\label{sce1}

\begin{figure}[ht]
\vskip -12mm
\centering
\includegraphics[scale=.4]{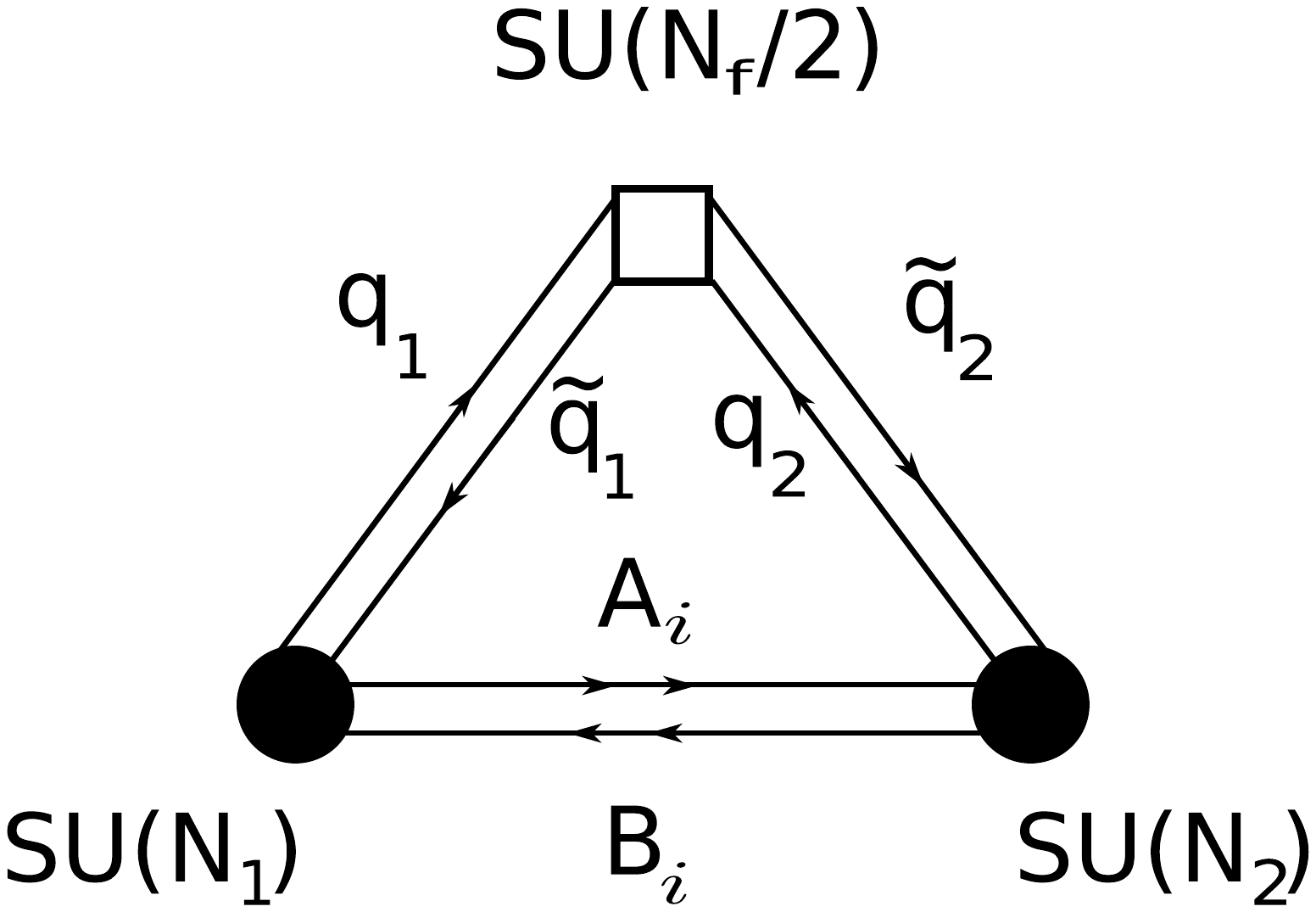}\vskip - 50mm
\caption{A Klebanov-Strassler quiver, flavoured by the addition of $N_f$ quarks.}
\label{shittyquiver}
\end{figure}

Using the fact that  the Klebanov-Strassler theory is on the 
baryonic 
branch, it is natural  to try to 
interpret the addition of infinitely extended D5 branes as true flavours. 
In particular, since the gauge coupling of the effective four dimensional 
theory on these branes is vanishing they should give rise to a  
global symmetry\footnote{This symmetry is typically broken to 
$U(1)^{N_f}$ due to the smearing.} $SU(N_f)_\mathrm{flavour}$.
We will then consider a quiver of Klebanov-Strassler type, with an explicit global flavour symmetry group and quarks transforming in fundamental 
representations of the gauge groups, as well as of the flavour symmetry group. In particular, 
the presence of a single stack of flavour D5 branes, together 
with the global $SU(2)\times SU(2)$ symmetry of the background, suggest to consider the following 
gauge and flavour groups\footnote{We may consider replacing $N_c\to N_c-N_f/2$, without altering the main points of the discussion.}
\bea
SU(N_c+n+n_f)\times SU(n+n_f) \times SU(N_f/2)_\mathrm{flavour},
\label{newprop}
\eea
and two pairs of quarks  $q_1,\tilde q_1$, and $q_2,\tilde q_2$, 
transforming under the two $SU(2)$ factors. The total number of quarks is 
then $N_f$.
Most of the discussion in this section applies with minor modifications to
a general class of quivers, where the number of quarks  of each gauge group may be different. In the following we will not specify 
a superpotential for the quarks. We will assume that there is a superpotential which allows to perform a sequence of 
Seiberg dualities.   The theory we are discussing is depicted in Figure \ref{shittyquiver}, where we 
denoted the ranks of the two gauge groups  generically with $N_1$ and $N_2$. 
Note that  this quiver is  in the same class  as the one discussed in 
\cite{Benini:2007gx} (cf. Figure 1 of this reference), where it was proposed to arise from a configuration 
with D7 branes. We study some aspects of this quiver  in Appendix \ref{quiverappendix}.

The first problematic issue that we face is that the ranks of the gauge  groups in (\ref{newprop}) 
include shifts  by the $n_f$  terms,  arising from the source D3 branes induced on the flavour
D5 branes, as we discussed earlier. Then the resulting theory does not have a baryonic branch \cite{Dymarsky:2005xt},
at least if the quarks do not have VEVs.  Turning on VEVs for the quarks 
might change this. However, in our solution $N_f$ is fixed, implying that 
the quarks cannot be Higgsed.  In any case, we have not found a solution to the D-terms and F-terms of this theory, 
with or without VEVs for the quarks, that may be interpreted as a baryonic branch. Moreover, the absence of the Goldstone boson mode in the gravity, 
gives support to the fact  that the theory is \emph{not} on a baryonic branch.

Building on the intuition gained from the discussion of the limiting case $N_f\to 0$, we would like to interpret the present solution
in terms of a cascade of Seiberg dualities and Higgsing. 
Then the second problem that we face is that we are not able to match gravity computations 
to field theory computations in  a natural way. The discussion around (\ref{truewarp})
 suggests that the running number of bulk D3 branes is now given by 
$n\propto k (N_c-N_f/2)$, with the cascade step changing logarithmically in the UV as $k\sim (N_c-N_f/2)\log r$. 
If we assume that we can perform Seiberg dualities on the quiver 
in an alternate fashion on the nodes, it turns out that this is \emph{not} the cascade pattern that the field theory follows. 
See Appendix \ref{quiverappendix} for  details.
Indeed, the difference in ranks, at each cascade step $k$ in the quiver is $N_c- k N_f/2$, thus it 
changes with the number of steps.
This does not seem to be possible in a string theory
solution where the number of fivebranes is fixed, and only the number of D3 branes varies. 
Moreover, the rate of change of the ranks goes like  $k^2N_f$, which is also not visible in our supergravity solution. 
One might speculate that there exists a particular pattern of Higgsing 
such that its contribution will exactly balance the terms of the cascade, in such  a way that the final result detected by the gravity solution is that in (\ref{truewarp}). To us this does not appear to be natural, nor likely, and it is not clear how one could check such a conjecture.

Finally, let us also mention that  the supergravity  beta functions 
obtained from identifying the dilaton with the sum of the gauge couplings, 
and the period of $B$ with the difference, do  not match with the field theory beta functions -- see Appendix \ref{quiverappendix}.
In particular, in the field theory  we have that the  beta function for the difference of 
gauge couplings is  $3N_c$, independently of $n,n_f,N_f$, which does not match the supergravity definition.
Of course, there is no proof  that these two calculations should agree, 
so this mismatch is not a rigorous 
objection to the validity of the present type of field theory scenario. 
We will discuss a little more beta functions in the following section, where we analyse a different field theory picture. 

In summary, if we follow the standard lore and 
add quarks transforming in the fundamental representation of the global flavour symmetry, 
the resulting quiver field theory does not seem to be consistent with the properties of our gravity solutions. 
At least, if we work in the hypothesis that we can interpret everything in terms of cascade plus Higgsing. 
This interpretation is satisfactory in the limit $N_f\to 0$, when the quarks 
in the quiver of Figure \ref{shittyquiver} are absent.

\subsubsection{Emerging flavour symmetry}
\label{sce2}

Assuming that a cascade of Seiberg dualties is taking place, we have seen
that in the previous field theory picture, these are not reflected by the supergravity behaviour. 
In addition, the field theory beta functions do not match to the usual supergravity definitions, as 
one would have naively expected. It is then natural to wonder whether there 
exists a different picture, in which  these features will 
match to the canonical 
expectations from the gravity side.  We have been able to realise this only assuming that the field
theory is a  two-node quiver of the Klebanov-Strassler type,  \emph{without} explicit flavours added. 
In particular, let us then consider the following gauge groups
\bea
SU(N_c+n+n_f)\times SU(N_f/2+n+n_f).
\label{kkktmba2}
\eea
After discussing how this proposal reproduces successfully various 
properties of the gravity solution, 
we will address what appears to be the main shortcoming of this picture, namely the role of the global flavour symmetry. 
Notice that this theory reduces correctly to the field theory discussed in section \ref{deformedzzz} 
in the limit $N_f\to 0$, at fixed $n_f$. Moreover, we can imagine removing all the $n+n_f$ D3 branes by undoing the ``rotation'' procedure.
The theory then reduces essentially to that proposed in 
\cite{Casero:2006pt,Casero:2007jj}, 
modulo the fact that the second group appears to be 
gauged. We will return to this point shortly.  

First of all, the cascade of Seiberg dualities here matches to the gravity calculations in the usual way. 
We can define a running number $n$ of bulk D3 branes by integrating 
$H_3\wedge F_3$ over the internal six-dimensional geometry (cut off at some distance in the UV), 
or equivalently from the corresponding UV term of the warp factor. Either 
way, one gets  $n \propto g_s (N_c-  \frac{N_f}{2})^2\log r$, 
and defining the cascade step $k$ from the period of $B$ over the two-sphere at large distance, we get
\bea
k \propto g_s (N_c-  \frac{N_f}{2}) \log r ~~~~\Rightarrow ~~~~ n\propto k (N_c-  \frac{N_f}{2}) 
\eea
Then more precisely our Klebanov-Strassler  quiver in (\ref{kkktmba2}) reads 
\bea
SU(N_c+k(N_c - N_f/2)+n_f)\times SU(N_f/2+k(N_c- N_f/2)+n_f).
\label{refinedquiver}
\eea
Of course the $n_f$ D3 branes may be still interpreted as being Higgsed, as we discussed. 
We now  \emph{assume} the validity of the following definitions for the 
gauge couplings  and theta angles
\cite{Klebanov:2000hb}: 
\bea
& & \frac{4\pi^2}{g_1^2}+ \frac{4\pi^2}{g_2^2}= \pi e^{-\phi},\;\;\;\;~~
\frac{4\pi^2}{g_1^2}- \frac{4\pi^2}{g_2^2}= 2\pi
e^{-\phi}(1-b_0),\nonumber\\
& & \Theta_1+\Theta_2= -2\pi C_0,\;\;\;\;~~~~
\Theta_1-\Theta_2=\frac{1}{\pi}\int_{\Sigma_2}C_2,
\label{couplingtheta}
\eea
where we defined 
\bea
b_0=\frac{1}{4\pi^2}\int_{\Sigma_2}B_2
\eea
and the integrals are performed on the two-cycle defined by 
$\Sigma_2=[\theta=\tilde{\theta},\varphi=2\pi-\tilde{\varphi}, \psi=\pi]$.
It is well known that these definitions can be justified rigorously only in very special cases, 
but nevertheless they do capture the correct field theory quantities 
(at least in the UV) in a variety of cases. 
Evaluating the beta functions using the standard energy-radius relation \cite{Klebanov:2000hb}
we then obtain (in the UV) 
\bea
\beta_1 = - \beta_2 = 3 (N_c - \frac{N_f}{2}),~~~~~\;\;\;\Delta \Theta_1=-\Delta \Theta_2= \alpha(2N_c-N_f)~.
\label{betasanomalies}
\eea 
Assigning (in the UV) anomalous dimensions and  
R-charges to the bi-fundamentals, exactly as done in \cite{Klebanov:2000hb}
\bea
\gamma_{A_i}=\gamma_{B_i}=-\frac{1}{2},\;\;\; R_{A_i}=R_{B_i}=\frac{1}{2}
\eea
we find that  the field theory calculations  reproduce the result in (\ref{betasanomalies}).
We note that the computation of the holographic central charge given in Appendix 
\ref{cchargeapp} is perfectly compatible with the quiver theory that we are discussing here.

The remaining issue to address is the fate of the expected $SU(N_f/2)$ global flavour  symmetry in this theory.  
We would like to propose that we can imagine the quarks arising in the IR,  
after the cascade of  Seiberg dualities and the Higgsing have taken place. 
In other words, these should be the bi-fundamentals $A_i,B_i$, in 
a configuration in which one gauge group is very weakly coupled, and can be 
thought of effectively as a flavour. More precisely, writing the matrix indeces 
explicitly (suppressing the $SU(2)\times SU(2)$ indeces), 
we can define $a^p{}_j = A^p{}_j$, $b^j{}_p = B^j{}_p$ and   $q^p{}_\alpha = A^p{}_\alpha$, $\tilde q^\alpha{}_p = B^\alpha{}_p$,  
where $p$ labels $\mathbf{N_c+ n+n_f}$, $j$ labels $\mathbf{n+n_f}$ and $\alpha$ labels $\mathbf{N_f/2}$.
At every cascade step $N_c$ and $N_f$ stay fixed while $n+n_f$ decreases, and similarly, at every Higgsing step the bi-fundamentals $a,b$
get mesonic type VEVs. Hence this splitting is well defined at any step. 
At the end of this process, the $a,b$ bi-fundamentals will disappear, leaving the $q,\tilde q$ ``quarks''. 
This is similar to the discussion in Klebanov-Strassler (see around eq.(109) in \cite{Klebanov:2000hb}), 
where the authors proposed that the addition of a small number $p$ of probe D3 branes at the bottom of the cascade 
should correspond to a $SU(M+p)$ gauge theory, with $2p$ ``flavours''. 
Indeed, if we take $N_f/2\ll N_c$, at the bottom of the cascade/Higgsing process,
the gauge coupling of the $SU(N_f/2)$  group will be much smaller than that of the $SU(N_c)$. 
Thus, at least in this case, this proposal is reasonable.

Let us make a few more comments in support of this idea. First of all recall that the global 
flavour symmetry in the dual theory arises from the gauge symmetry on the world-volume of the flavour D5 
branes. In particular, global currents are given by  fluctuations of the gauge fields on the branes. However, if 
there were  {\it no} normalisable fluctuations, 
then we would not have any state charged under this symmetry, and the 
global symmetry would be effectively
 gauged\footnote{Notice that {\it before} the ``rotation'' the $N_c$ D5-$N_f$ 
D5 solution is not decoupled from gravity,  hence any  putative global symmetry should be gauged. Only in the $c\to 0$ limit does the 
solution correspond to a field theory with $N_f$ flavours.}.
Given our asymptotics, it may be the case that all fluctuations of the gauge fields on the flavour branes are 
non-normalisable.  It would be interesting to prove or disprove this statement.
We also note that in our solution the dilaton $e^{-\Phi}$ is divergent at the origin (see 
(\ref{thedilat})), suggesting one of the gauge couplings vanishes in the IR. 
We do not know whether we can trust this, or if it is an artifact of the 
IR singularity.  Finally, recall that in the ordinary cascade one can get 
close to points where one 
of the couplings vanishes - see \cite{Strassler:2005qs} - and we can think 
that the Higgsing may eventually freeze the running at this point.

Interestingly, note that in the case $N_f=2N_c$ the quiver 
(\ref{refinedquiver})  becomes actually  a Klebanov-Witten \cite{Klebanov:1998hh} quiver
\bea
SU(N_c+n_f)\times SU(N_c+n_f).
\label{KWquiver}
\eea
It is satisfying that  in this case the gravity calculations show that  
the beta functions and the anomalies vanish in the UV. 
Moreover, the cascading behaviour disappears from the solution, 
as can be seen for example from the leading behaviour of the warp factor.
This solution may be then describing a Klebanov-Witten theory, 
where conformal invariance is broken by the Higgsing effect, 
represented by the leading term in the warp factor. 
As can be seen neatly from  the presentation in Appendix \ref{mmvar}, 
in this case a $\Z_2$ symmetry is restored in the solution.
The $U(1)_R$ (corresponding to $\psi \to \psi + \delta$) is broken to $\Z_2$ but, differently  
from \cite{Klebanov:2000nc,Klebanov:2000hb},  it is not broken 
in the UV, as expected from the fact that this is not anomalous in this case.

We note that there exists also an analytic solution, which \emph{does not} 
correspond to setting $N_f=2N_c$ in the solution discussed in the rest of 
this paper. In this solution the $U(1)_R$ is preserved, 
while there is no $\Z_2$ symmetry. We briefly discuss this solution in Appendix \ref{ananal}. 
Solutions with $N_f=2N_c$ and their possible desingularisations are 
clearly interesting in their own right.
We leave further studies of these solutions for future work.

\vskip 5mm

The two scenarios illustrated above are by no means exhaustive and perhaps 
a more sophisticated set-up is necessary to obtain a better understanding 
of the field theory  dual to the background we have presented (if indeed 
this exists).

\section{The $\Z_2$ in gravity and in field theory}
\label{thez2}

In this section we will discuss a $\Z_2$ symmetry in gravity, as well as its  field theory interpretation.

\subsection{Back to gravity}

\begin{figure}[h]
\centering
\includegraphics[scale=.9]{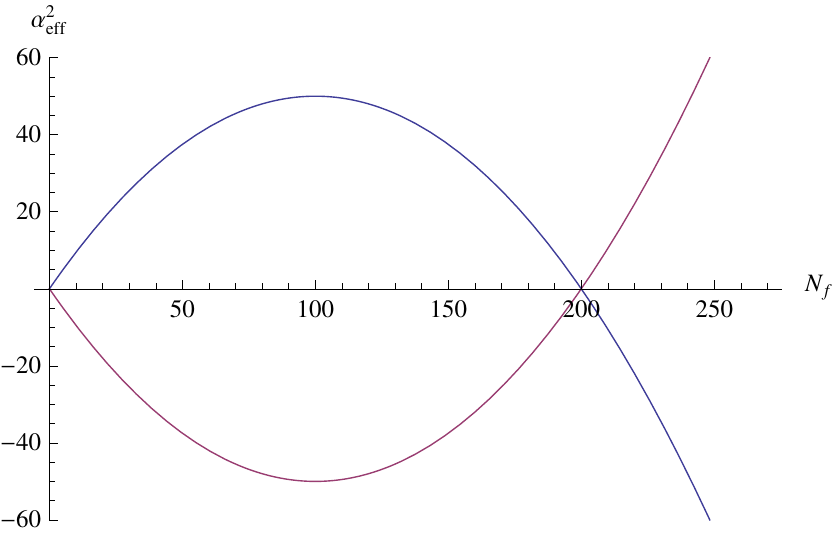}
\caption{Plot of the effective resolution parameter $\alpha^2_{eff}$ as a function of $N_f$ at fixed $\nu$ and (large) $\rho$. The two branches are exchanged by a $\Z_2$ reflection. At $N_f=0$ the $\Z_2$ symmetry ${\cal I}$ is restored. At $N_f=2N_c$ the reflection symmetry ${\cal R}$ is restored.}
\label{effectiveres}
\end{figure}

Following \cite{Maldacena:2009mw}, we can define  an effective (running) 
resolution parameter $\alpha^2_{eff}$ as the size of the two-sphere measured from infinity. 
In the unwarped metric this gives $\alpha^2_{eff} \sim (N_c-N_f/2)\rho $. See Appendix \ref{mmvar}. In the warped metric after the rotation, this becomes multiplied by the parameter $c^{-1}$, hence we can write this as 
\bea
\alpha^2_{eff} \sim \frac{N_f}{\nu}(N_c-N_f/2)\rho .
\eea
We can then imagine starting from the solution of section 
\ref{flavKS} and increase $N_f$ from zero, keeping $\nu$ fixed. Interestingly, the effective 
resolution will then increase, reaching a maximum at $N_f=N_c$ and then will vanish again at $N_f=2N_c$. 
Increasing $N_f$ further, the resolution 
parameter will become negative. There is another solution, obtained by a 
$\Z_2$ symmetry, which in Figure 
\ref{effectiveres} corresponds to reflection along the horizontal axis. The points $N_f=0$ and $N_f=2N_c$ are special, as we now discuss. 

In the gravity side there is  the $\Z_2$ action which  interchanges the two-spheres defined as 
\bea
{\cal R}: 
\theta \leftrightarrow \tilde \theta, ~~~~ 
\varphi \leftrightarrow \tilde \varphi.
\eea
This can be combined with the 
change of sign of the three-forms $H_3 \to -H_3,  F_3 \to - F_3$   (the center of $SL(2,\Z)$) to define the following 
$\Z_2$ action
\bea
{\cal I}: ~~\theta \leftrightarrow \tilde \theta, ~~~~ \varphi \leftrightarrow \tilde \varphi, ~~~~ H_3 \to -H_3, ~~~~ F_3 \to - F_3.
\eea
Notice that the effective resolution parameter $\alpha^2_{eff}$ changes sign under ${\cal R}$ and ${\cal I}$.
These operations have been discussed in the context of the Klebanov-Witten 
\cite{Klebanov:1998hh} and Klebanov-Strassler  \cite{Gubser:2004qj} theories. 
Let us briefly summarize the behaviour of various gravity solutions under the ${\cal I}$ (and ${\cal R}$) action:

\begin{itemize}

\item 
Unflavoured. Unrotated. For any value of $c<\infty$ the effective resolution is non-zero, hence the metric is not invariant, 
while the $F_3$ is invariant. 
Therefore the full solution is \emph{not invariant} 
under ${\cal I}$ (neither invariant under ${\cal R}$).

\item 
Unflavoured. Rotated. For any value of $c<\infty$ the effective resolution is non-zero, hence the metric is not 
invariant,  while the $F_3$ is invariant. 
Therefore the full solution is \emph{not invariant} under ${\cal I}$ (and ${\cal R}$).  
For $c=\infty$ we have the Klebanov-Strassler solution: the effective resolution vanishes hence 
the metric becomes invariant. The $F_3$ is invariant. Hence the full solution is \emph{invariant} under ${\cal I}$ 
(but not under  ${\cal R}$).

\item 
Flavoured. Unrotated. For any value of $c$ and $N_f\neq 2N_c$ 
the effective resolution is non-zero, hence the metric is not 
invariant. The $F_3$ is not invariant. 
Hence the full solution is \emph{not invariant} under ${\cal I}$ (and ${\cal R}$). 
If we exclude the solutions in Appendix \ref{ananal}, for any value of $c$ 
and 
$N_f= 
2N_c$ the effective resolution vanishes hence the 
metric is invariant.  The $F_3$ is not invariant under ${\cal I}$ but it is invariant under ${\cal R}$. 
Hence the full solution is \emph{not invariant} under ${\cal I}$ but it is \emph{invariant} under ${\cal R}$.

\item 
Flavoured. Rotated. For any value of $c<\infty$ and $N_f\neq 2N_c$ 
the effective resolution is non-zero, hence the metric is not, and the $F_3$ is not invariant. 
Hence the full solution is \emph{not invariant} under ${\cal I}$ (and ${\cal R}$).  
For any value of $c$ and $N_f= 2N_c$ the metric is invariant. The $F_3$ is not invariant is 
 under ${\cal I}$ but it is invariant  under ${\cal R}$. Hence the full solution is \emph{not invariant} under ${\cal I}$ but it is \emph{invariant} under ${\cal R}$. 
 For $c\to\infty$ and $N_f\to0$ the effective resolution vanishes hence the metric is invariant, 
and the $F_3$ is invariant. Hence the full solution is \emph{invariant} under ${\cal I}$ (but not under ${\cal R}$).

\end{itemize}

\subsection{Field theory}

In the Klebanov-Strassler field theory it is argued in the literature that  ${\cal I}$ acts as follows:
\bea
{\cal I}: ~~ A_i \to B_i^\dagger, ~~~~ B_i \to A_i^\dagger 
\eea
which is interpreted as simultaneous exchange of $A_i$ and $B_i$, 
accompanied by charge conjugation (without exchange of gauge groups, which does not make sense for unequal ranks).
Recalling that  the Klebanov-Strassler 
superpotential is  $W = Tr[A_1B_1A_2B_2]-Tr[A_1B_2A_2B_1]$ we see that 
\bea
{\cal I}: ~~ W \to  W^*.
\eea
Since we have to take the real part of $W$ in the Lagrangian, we see that this is invariant under ${\cal I}$. 
On the other hand, we have 
\bea
{\cal I}:~~ {\cal U} = \sum_iTr[A_iA_i^\dagger  - B_i^\dagger B_i] \to - {\cal U},
\eea
which explains why the Klebanov-Strassler point does not break 
${\cal I}$, while for generic points on the baryonic branch, this is broken.

We now want to study the action of ${\cal I}$ when there are quarks, coming from flavour branes. 
Notice that under ${\cal I}$
the smearing form $\Xi_{(4)} $ changes sign. This means that the flavour fivebranes change orientation, 
and suggests that we should still apply charge conjugation on the quarks. 
The interchange of the bi-fundamentals $A_i$ and $B_i$ under ${\cal I}$ 
is dictated by the fact that the interchange of the two-spheres is a ``geometrical'' operation 
\cite{Klebanov:1998hh}. To see this, it is convenient to 
write the (resolved) conifold equation\footnote{This calculation is justified by thinking about the brane
set up before the back-reaction.} as $w_1^2 + w^2_2 + w_3^2 + w_4^2 = 0$. 
By writing the $w_i$ coordinates  explicitly in terms of angular 
coordinates it is then easy to verify  that \cite{Klebanov:1998hh}
\bea
{\cal I}: ~~ w_1 \to w_1, ~~~~  w_2 \to w_2, ~~~~  w_3 \to w_3, ~~~~ w_4 \to - w_4 .
\eea
Then a further change to the variables $z_i$  defined as 
$w_1 = z_1 + z_2, i w_2 = z_1 - z_2,   w_3 = z_3 + z_4,  i w_4 = z_3 - z_4$, where 
$z_1 = A_1B_1, z_2 = A_2B_2,  z_3 = A_1 B_2,   z_4 = A_2 B_1$, makes clear why ${\cal I}$ interchanges $A_i$ with $B_i$.
On the other hand, the quark fields come from the open string stretching between the flavour fivebranes, and they are not 
related to the geometry. It is therefore plausible that the quarks should not be interchanged under ${\cal I}$.  
Then generically, if we imagine that our field theory has bi-fundamentals $a_i,b_i$ and quarks
 $q_i,\tilde q_i$, we can postulate the following action 
\bea
{\cal I}: ~~ a_i \to b_i^\dagger, ~~~~ b_i \to a_i^\dagger, ~~~~ q_i \to q_i^\dagger, ~~~~ \tilde q_i \to \tilde q_i^\dagger .
\eea

In order to see whether this is a symmetry of a field theory, we need to specify the Lagrangian and the vacuum.  
Let us concentrate in the field theory scenario discussed in \ref{sce2}.
As explained earlier, we are thinking that the ``quarks'' are part of the bi-fundamental fields. 
In matrix notation, the splitting discussed in section \ref{sce2} can be written as 
\bea
A = (a ~ q), ~~~~~~~~B^T = (b^T ~\tilde q^T),
 \label{splitt}
\eea
where $a, b^T$ are $(N_c+n+n_f)\times (n+n_f)$ matrices and $q,\tilde q^T$ are  $(N_c+n+n_f)\times N_f/2$ matrices.  
When we write these expressions we are making manifest only a $SU(N_f/2)\times SU(n+n_f)$ sub-group 
of the  $SU(N_f/2+ n+n_f)$ gauge symmetry. 
If we apply  $SU(N_f/2+ n+n_f)$ rotations, the $a$ and $q$, and $b$
and $\tilde q$ in (\ref{splitt}) will mix. However, it is legitimate to define how a symmetry (namely ${\cal I}$) 
acts in a convenient gauge. Notice also  that in our picture by definition the $a,b$ will get mesonic-type VEVs during the 
Higgsing process, while the $q,\tilde q$ will not get VEVs. This gauge choice is then also convenient for this purpose, 
since as we cascade/Higgs, the rank of $SU(n+n_f)$ will decrease, while $SU(N_f/2)$ will stay fixed. 

Writing the Klebanov-Strassler quartic superpotential  in terms of $a,b,q,\tilde q$ we have:
\bea
W = W_1 + W_2 + W_3 
\eea
where 
\bea
W_1 & =& Tr[a_1b_1a_2b_2]-Tr[a_1b_2a_2b_1]\nonumber\\
W_2 &= & Tr[q_1\tilde q_1 q_2 \tilde q_2]-Tr[q_1 \tilde q_2 q_2 \tilde q_1]\nonumber\\
W_3 & = &  Tr[q_1\tilde q_1 a_2b_2] + Tr[a_1b_1 q_2\tilde q_2]  - Tr[q_1\tilde q_2 a_2b_1] - Tr[a_1b_2 q_2\tilde q_1].
\eea
Then we have (the real part is understood) 
\bea
{\cal I }: ~~ W_1 \to W_1,~~~~ W_2 \to - W_2 
\eea
and 
\bea
{\cal I }: ~~ W_3 \to W_3'= Tr[\tilde q_1 q_1 a_2b_2] + Tr[a_1b_1 \tilde q_2 q_2]  - Tr[\tilde q_1 q_2 a_2b_1] - Tr[a_1b_2 \tilde q_2 q_1].
\eea
Notice that the transformed $W_3'$  is not related simply to the initial $W_3$.
However, this still makes sense, since in the transformed theory we reversed the arrows on the quarks.

We can ask whether the mesonic VEVs turned on in the Higgsing process
can break the $\Z_2$. In fact, if we 
think of adding D3 branes at points in the deformed conifold, in general the resulting solution 
will break all the symmetries of the deformed conifold. 
However, since in our set up the source D3 branes are smeared uniformly, 
all the symmetries that would have been broken at a generic point on 
the mesonic branch, are restored. This is obvious from the fact that our limiting 
Klebanov-Strassler-like solution has all the symmetry of the deformed conifold, including the $\Z_2$. 
Hence in our solution we should think of the $\Z_2$ breaking as due only 
to the quarks, and not to the mesonic VEVs.

As we discussed, in the case $N_f=2N_c$  the metric \emph{and} the three-forms become invariant under the interchange of two-spheres,
hence a different $\Z_2$ symmetry appears. Since now the ranks of the gauge groups are equal, it seems natural to assume that this 
$\Z_2$ action in the field theory corresponds to the interchange of $A_i$ and $B_i$, together with the exchange of the two gauge groups, as discussed 
in \cite{Klebanov:1998hh}\footnote{This operation changes the overall sign of the superpotential. This may be reabsorbed by the symmetry $\Upsilon$, which by definition leaves invariant the lowest components of the superfields $A_i, B_i$ \cite{Klebanov:1998hh}.}. 
In our case this operation exchanges the role of ``flavours'' and ``colours'', which of course can make  sense only in the quiver  (\ref{KWquiver}).

\section{Discussion}\label{sectiondiscussionzz}

In this paper we have used a solution generating transformation, 
applicable to a  large class of supersymmetric Type IIB backgrounds, to construct a new family of 
solutions generalising the resolved deformed conifold of Butti \emph{et al} \cite{Butti:2004pk}.  Using this method, we can take any solution to the torsional 
superstring equations \cite{Strominger:1986uh,Hull:1986kz,GMW}  and generate
a solution where various RR and NS fields are turned on. 
The method may be applied to solutions which include the back-reaction  of smeared 
source branes, usually referred to as flavour branes. 
In particular, we have applied this procedure to a solution 
representing a system of $N_c$ D5 branes 
wrapped on the two-sphere inside the resolved conifold, 
with addition of $N_f$ flavour D5 branes, wrapped on a transverse infinitely extended cylinder.
The final solution  is then a warped resolved deformed conifold, modified by the back-reaction of the extra flavour branes.

The flavoured solution differs qualitatively from the unflavoured one in two ways. Firstly, it is singular in the IR. Secondly, 
the UV asymptotics is not the (logarithmic) Klebanov-Strassler one. 
Although we have not addressed the resolution of the IR singularity in this paper, 
we expect it may be resolved by considering
a profile for the flavour D5 branes that vanishes smoothly in the IR. 
The different behaviour in the
UV is induced by the presence of a uniform distribution of source D3 branes, smeared on the transverse geometry, up to infinity.
We have explained
that these D3 branes are induced by the presence of the flavour D5 branes
in a  geometry with a non-trivial $B$-field.  In other words, the ``rotation'' procedure has the effect of adding bulk D3 branes, coming from the original colour D5 branes, and smeared source  D3 branes, coming from the original flavour D5 branes.

This geometric set-up leads to an unusual picture in the dual field theory. First of all,  
the different asymptotic behaviour of the warp factor implies that the zero 
mode associated with changing the  parameter $c^{-1}$ is here \emph{non-normalisable}. 
This suggests that our gravity solutions correspond to vacua where the global baryonic $U(1)_B$ is not 
spontaneously broken. In particular, assuming that the field theories are some modifications of the 
Klebanov-Strassler theory, we have proposed that our  
solutions correspond to \emph{mesonic branches}. 
This matches nicely with the fact that we always have $n_f$ extra D3 branes in the geometry, which are running because of a 
\emph{Higgsing} effect.  We can then explain the  parameter $c^{-1}$ purely as 
arising from the back-reaction of the $N_f$ flavour D5 branes in the geometry. 
In particular, since changing $c^{-1}$ corresponds to changining $N_f$ (at fixed $n_f$), this is a change of theory.
This is  rather different  from the interpretation in the case without any source D3 or D5 branes, where this
parameter is related to the classical VEV of the operator ${\cal U}$, which is the partner of the baryonic $U(1)_B$ current. 
This different interpretation of a background 
in the presence of extra sources should not be too surprising. Indeed, 
the Klebanov-Strassler geometry, plus  some probe D3 branes is indeed dual to the field theory in the mesonic branch.  
On  the mesonic branch we do not expect a fuzzy two-sphere emerging from the field theory.  
The asymptotic growth of the warp factor in our solution indicates that the effective radius of the two-sphere in the 
closed string metric is never much smaller than the $B$ field \cite{Maldacena:2009mw}, 
suggesting that a fuzzy sphere interpretation is in this case not applicable \cite{Seiberg:1999vs}.  
However, we have not checked this in detail.

The general picture that we have proposed is that in our solutions the decrease of five-form flux, 
and of ranks in the gauge theories, is due 
to a sequence of steps involving both Seiberg dualities and Higgsing. In the case where we turn off the $N_f$ 
flavour D5 branes (but not the $n_f$ D3 sources) this picture is robust. 
Turning on $N_f$ led us to discuss  slightly more exotic scenarios.  
The standard lore about ``flavouring'' field theories instructs us to pick an unflavoured theory and 
add quarks transforming under a new global flavour symmetry. However, we have seen that in the present case
this recipe leads to a field theory that fails to reproduce the main properties
of the gravity side. In particular, it does not seem possible to reproduce the pattern of Seiberg dualities.  
This is perhaps not very surprising, given that flavoured versions of the Klebanov-Strassler theory
of the type we considered in section \ref{sce1}
actually were shown to arise from constructions involving D7 branes  \cite{Benini:2007gx}.
We have then analysed the possibility that the flavour symmetry might actually emerge only in the IR. Although 
this may be an unorthodox suggestion, we have explained that such proposal 
matches all the expected gravity 
predictions. However, checking this statement in our solution is difficult, since this is singular in the IR.

We have indicated  that the singularity in the IR may be resolved. Hopefully further studies of non-singular versions 
of our solutions  will clarify the validity of the ideas presented in this paper.

\section*{Acknowledgements}

We would like to  thank O. Aharony, J. Maldacena, D. Rodriguez-Gomez, 
and J. Sparks for discussions and useful  comments. Special thanks to Ofer 
Aharony for detailed discussions.

\appendix
\renewcommand{\thesection}{\Alph{section}}
\renewcommand{\theequation}{\Alph{section}.\arabic{equation}}

\section*{Appendices}

\section{Type IIB equations of motion with sources}
\setcounter{equation}{0}

In this appendix  we state the equations of motion coming from the action \eqref{eq:flavoured_action}. 
Let us first rewrite the total action, comprising the Type IIB supergravity action plus  the supersymmetric source action:
\be
	S = S_{IIB} + S_{\text{sources}}
\ee
where
\be
	\begin{aligned}
		S_{IIB} = &\int \sqrt{-g} \left( R - \frac{1}{2} \partial_{\mu} \Phi \partial^{\mu} \Phi \right) + \frac{1}{2} \int C_{(4)} \wedge F_{(3)} \wedge H\\
		&- \frac{1}{2} \int \left( e^{2\Phi} F_{(1)} \wedge *F_{(1)} + e^{-\Phi} H \wedge *H + e^{\Phi} F_{(3)} \wedge *F_{(3)} + \frac{1}{2} F_{(5)} \wedge *F_{(5)} \right)
	\end{aligned}
\ee
and
\be
	S_{\text{sources}} = - \int \Big(e^{4\Delta + \Phi/2} \vol_{(4)} \wedge \big( \cos \zeta e^{2\Delta} J + \sin \zeta e^{-\Phi/2} B \big) - C_{(6)} + C_{(4)} \wedge B \Big) \wedge \Xi_{(4)}
\ee
In the following we will set $F_{(1)} =0$. The modified Bianchi identities for the fluxes read
\be
	\begin{aligned}
		\dd H &= 0 \\
		\dd F_{(3)} &= \X_{(4)} \\
		\dd F_{(5)} &= H \wedge F_{(3)} + B \wedge \X_{(4)} \\
	\end{aligned}
\ee
The equations of motion for the fluxes read
\be
	\begin{aligned}
		\dd \left( e^{-\F} * H \right) &= F_{(3)} \wedge F_{(5)} + \sin \z e^{4\D} \vol_{(4)} \wedge \X_{(4)} \\
		\dd \left( e^{\F} * F_{(3)} \right) &= - H \wedge F_{(5)}
	\end{aligned}
\ee
Notice that in the equation of motion for $H$ the term coming from $C_{(4)} \wedge B \wedge \X_{(4)}$ in the source action is exactly cancelled by a contribution from the Chern-Simons term of the Type IIB supergravity equations. 
We then define the following notation
\be
	\o_{(p)} \lrcorner \l_{(p)} = \frac{1}{p!} \o^{\mu_1 ... \mu_p} \l_{\mu_1 ... \mu_p}
\ee
We also have that
\be
	\int \o_{(p)} \wedge \l_{(10-p)} = - \int \sqrt{-g} \l \lrcorner (*\o)
\ee
Using these, we can write the dilaton equation of motion as
\be
	\frac{1}{\sqrt{-g}} \partial_{\mu} \left( \sqrt{-g} g^{\mu \nu} \partial_{\nu} \F \right) = 
\frac{1}{12} e^{\F} F_{(3)}^2 - \frac{1}{12} e^{-\F} H^2 - \frac{1}{2} e^{\F/2} \X_{(4)} \lrcorner * \left( \cos \z e^{6\D} \vol_{(4)} \wedge J \right)
\ee
Finally, the Einstein equation is
\be
	\begin{aligned}
		R_{\m \n} = &\frac{1}{2} \partial_{\m} \F \partial_{\n} \F + \frac{1}{48} e^{\F} \left( 12 F_{\m \r \s} F_{\n}^{\phantom{\n} \r \s} - g_{\m \n} F_{(3)}^2 \right) +	\frac{1}{48} e^{-\F} \left( 12 H_{\m \r \s} H_{\n}^{\phantom{\n} \r \s} - g_{\m \n} H^2 \right) \\
		&+ \frac{1}{96} F_{\m \r_1 \r_2 \r_3 \r_4} F_{\n}^{\phantom{\n} \r_1 \r_2 \r_3 \r_4} \\
		&- \frac{1}{24} e^{6\D+ \F/2} \cos \z \left( 2 \X_{\m \r_1 \r_2 \r_3} * \left( \vol_{(4)} \wedge J \right)_{\n}^{\phantom{\n} \r_1 \r_2 \r_3} -3 g_{\m \n} \X_{(4)} \lrcorner * \left( \vol_{(4)} \wedge J \right) \right) \\
		&- \frac{1}{240} \sin \z e^{4\D} \left( (B \wedge \X_{(4)})_{\m \r_1 ... \r_5}  \left( * \vol_{(4)} \right)_{\n}^{\phantom{\n} \r_1 ... \r_5} - 60 g_{\m \n} (B \wedge \X_{(4)}) \lrcorner \left( * \vol_{(4)} \right) \right)
	\end{aligned}
\ee

\section{More on the solutions}
\setcounter{equation}{0}
\label{more-on-solutions}

In this appendix we discuss some properties of the master equation (\ref{master}) and we 
present an analysis of the asymptotic form of the solutions discussed in the main body of the paper. 
Moreover, we desribe a systematic method for constructing these solutions recursively.   
Let us start by noticing that the master equation (\ref{master}) can be written in the compact form 
\be\label{master-transformed}
P=s\pa_\r\left(\frac{P^2-Q^2}{4s(P'+N_f)}\right),
\ee
where
\be
s=\frac{1}{\sinh^2(2(\r-\r_o))}.
\ee     
Defining 
\be
\cp\equiv 2\int d\r s^{-1}P,
\ee
we obtain
\be
\cp s\pa_\r(s\pa_\r\cp)-\frac14(s\pa_\r\cp)^2+2N_fs\cp+Q^2=0.
\ee
Moreover, introducing the variable
\be
t\equiv 2\int d\r s^{-1}=\left\{ \begin{matrix}
                                 \frac12 e^{4\r}, & \r_o\to-\infty,\\
				 \frac14(\sinh(4(\r-\r_o))-4(\r-\r_o)), & \r_o>-\infty,
                                \end{matrix}\right.
\ee
we have $P=\dot{\cp}$ and the master equation takes the form
\be\label{master-simple}
4\cp\ddot{\cp}-\dot{\cp}^2+g_1(t)\cp+g_0(t)=0,
\ee
where the dots denote derivatives with respect to
 $t$ and the functions $g_1$ and $g_2$ are given by
\be
g_1(t)= 2N_f s,\quad g_0(t)= Q^2.
\ee

Equation (\ref{master-simple}) is equivalent to (\ref{master}), but it is only quadratic in $\cp$, which considerably
simplifies the systematic analysis of the equation. On the other hand, for 
$\r_o>-\infty$ the functions $g_1(t)$ and $g_0(t)$ are only parametrically known in terms of $\r$, but this will not be a
serious obstacle in the following analysis. Notice also that (\ref{master-simple}) is manifestly
a generalisation of eq. (A.6) of \cite{Maldacena:2009mw} to the case with 
$N_f\neq 0$, upon identifying
$\cp=N_c f/4$ and $t=\t$.  
The simplified master equation (\ref{master-simple}) can be derived from the simple action
\be
S=\int dt \cp^{-1/2}\left(\dot{\cp}^2-g_1(t)\cp+g_0(t)\right),
\ee
which again is a generalisation of the Lagrangian (A.7) of 
\cite{Maldacena:2009mw} for $N_f\neq 0$. The 
corresponding Hamiltonian is given by
\be\label{effective-hamiltonian}
H=\cp^{-1/2}\left(\frac14 \cp \p^2_\cp+g_1(t)\cp-g_0(t)\right),
\ee 
where the canonical momentum conjugate to $\cp$ is
\be
\p_\cp=2\cp^{-1/2}\dot{\cp}.
\ee
It follows that the solutions can be obtained from the Hamilton-Jacobi equation
\be\label{HJ}
\frac14\cp\left(\frac{\pa \cs}{\pa\cp}\right)^2+g_1(t)\cp-g_0(t)+\cp^{1/2}\frac{\pa\cs}{\pa t}=0,
\ee
with
\be
\p_\cp=\frac{\pa\cs}{\pa\cp}.
\ee
Obtaining a solution $\cs(\cp,t)$ of (\ref{HJ}) leads to a first order equation for $\cp(t)$ via the identification
\be
2\cp^{-1/2}\dot{\cp}=\frac{\pa\cs}{\pa\cp}.
\ee

\subsection{Recursive construction of the solutions}

It is known \cite{Casero:2006pt,Casero:2007jj,HoyosBadajoz:2008fw} that the solutions of (\ref{master}) 
fall into two possible categories according to their UV behaviour: they either lead (via (\ref{BPS-sols})) to an 
asymptotically linear or asymptotically constant dilaton. 
Since the ``rotation'' formulas (\ref{rotated-solution}) require
that the dilaton is bounded from above, it is only the solutions with the 
asymptotically constant dilaton that
can be rotated. As we have seen, these solutions
 behave as $P\sim c e^{4\r/3}$ in the UV, for some {\em strictly positive} 
constant $c$. In this section we present a systematic way of constructing all solutions with these UV asymptotics 
in an expansion for large $c$ \cite{HoyosBadajoz:2008fw}.     

Instead of constructing these solutions by expanding $P$, as was done in 
\cite{HoyosBadajoz:2008fw}, we will use (\ref{master-simple}) and write
\be
\cp(t)=\tilde{c} p(t),
\ee
where $\tilde{c}\equiv 3c/2$, has been chosen such that (for $\r_o=0$)
\be
\cp(t)\sim \tilde{c} t^{4/3}\Leftrightarrow P(\r)\sim c e^{4\r/3}.
\ee 
Expanding 
\be\label{c-expansion}
p(t)=\sum_{n=0}^\infty \tilde{c}^{-n}p_n(t),
\ee
and using (\ref{master-simple}) we get the sequence of equations
\bea
&&4p_0\ddot{p}_0-\dot{p}_0^2=0,\\
&&2p_0\ddot{p}_n-\dot{p}_0\dot{p}_n+2\ddot{p}_0p_n=-\frac12\left(g_1p_{n-1}+g_0\d_{n2}
+\sum_{m=1}^{n-1}\left(4p_m\ddot{p}_{n-m}-\dot{p}_m\dot{p}_{n-m}\right)\right),\quad n>0.\NO
\eea
The general solution of the equation for $p_0$ is
\be
p_0=a (t-t_0)^{4/3},
\ee
where $a$ and $t_0$ are constants. $a$ can be absorbed in $\tilde{c}$ and we choose $t_0=0$ so that $t\to 0$ as $\r\to \r_o$.
Using then $p_0=t^{4/3}$ in the rest of the equations, we find that the general solution is given by
\be\label{large-c}
p_n(t)=t^{4/3}\int_{t^+_n}^t dt' t'^{-7/3}R_n(t')
-t^{1/3}\int_{t^-_n}^t dt' t'^{-4/3}R_n(t'),\quad n>0,
\ee
where
\be
R_n(t)=-\frac14t^{2/3}\left(g_1(t)p_{n-1}+g_0(t)\d_{n2}
+\sum_{m=1}^{n-1}\left(4p_m\ddot{p}_{n-m}-\dot{p}_m\dot{p}_{n-m}\right)\right),
\ee
and $t_n^\pm$ are arbitrary constants. 

Some comments are in order here:
\begin{itemize}

\item Given this expansion for $\cp(t)$, $P(t)$ can be obtained by term-wise differentiation as
\be
P=\dot{\cp}=\sum_{n=0}^\infty \tilde{c}^{1-n}\dot{p}_n.
\ee

\item By setting $t_n^+=\infty$ one ensures that $p_n=o(t^{4/3})$ as $t\to\infty$, i.e. that $p_n$ for $n>0$ is 
asymptotically subleading in the UV compared to $p_0$. In fact, one can prove a stronger statement, namely that
the expansion (\ref{c-expansion}) is also an asymptotic expansion for the solutions that asymptote to the
deformed conifold as $t\to\infty$. In particular, with the choice $t_n^+=\infty$ and using the fact that 
\be
g_1(t)=\frac{N_f}{t}+\cdots, \quad g_0(t)=\frac{1}{16}\left(2N_c-N_f\right)^2\log^2(8t)+\cdots,
\ee
one can show iteratively that
\be
p_n(t)=\co\left(t^{(4-n)/3}(\log t)^{k(n)}\right),\; \text{as} \;t\to\infty,
\ee
where $k(n)$ is a positive integer. Hence, $p_{n+1}$ is asymptotically subleading relative to
$p_{n}$, i.e. not just compared to $p_0$. 

\item In the flavour-less case the expansion (\ref{c-expansion}) is {\em uniformly} valid for all $t$. In particular,
with a suitable choice for $t_n^-$ it correctly reproduces the IR asymptotic behaviour of the Butti et al.
solution \cite{Butti:2004pk}, as well as the UV behaviour. As we demonstrate below, this allows us to 
determine the constant that parameterises the IR asymptotics in an 
expansion in large $c$, the parameter that governs the
UV asymtptotics.    

\item Although the integrals involved in the expressions for $p_n(t)$, $n>0$, cannot be evaluated 
explicitly, the expansion (\ref{c-expansion}) is converging rather fast as we will see in the plots below. 

\item Inserting the expansion (\ref{c-expansion}) in the expression for the dilaton in (\ref{BPS-sols}) we obtain
\bea\label{dilaton-expansion}
e^{-2\F}&=&e^{-2\F_\infty}\left(1+\frac{1}{c}\pa_t(t^{-1/3}p_1)
+\frac{3}{2c^2}\left(\pa_t\left(t^{-1/3}p_2-\frac18t^{-5/3}p_1^2\right)\right.\right.\NO\\
&&\left.\left.+\frac38\left(\pa_t\left(t^{-1/3}p_1\right)\right)^2-\frac38t^{-2/3}g_0\right)+\co(1/c^3)\right),
\eea
where
\be
e^{-2\F_\infty}\equiv \frac49 e^{-2\F_o}\tilde{c}^{3/2}.
\ee
This formula is useful for evaluating the warp factor $\hat{h}$. 

\end{itemize}

\subsection{The unflavoured solution}

Let us now examine in more detail the expansion (\ref{c-expansion}) in the case $N_f=0$. Firstly, since
$g_1(t)=0$ in this case, it follows by induction that all odd terms $p_{2k+1}(t)$ vanish. Moreover, 
the unique value of $t_{2k}^-$ for which the expansion (\ref{c-expansion}) leads to a regular
solution as $t\to 0$ is $t_{2k}^-=0$. With this choice of the integration constants $t_{2k}^\pm$
the expansion (\ref{c-expansion}) for the flavour-less case is uniformly valid for all values of $t\in \mathbb{R}^+$ and
so one can immediately extract both the UV and IR behaviour of the solution. 

\begin{flushleft}
{\em UV asymptotics}
\end{flushleft}

As $t\to \infty$ we have
\be
g_0(t)=\frac{N_c^2}{4}\left(\log^2(8t)-4\log(8t)+4+\co\left(\frac{\log^2t}{t}\right)\right).
\ee
Using the recursion relations (\ref{large-c}) then one computes
\be
p_2(t)=\frac{9}{32}t^{2/3}\left(\log^2(8t)-7\log(8t)+\frac{47}{2}\right)+\co(t^{-1/3}\log^2(t)).
\ee
Differentiating this with respect to $t$ and using the relation
\be
t=\frac18 e^{4\r}-\r+\co(e^{-4\r}),\; {\rm as}\; \r\to\infty,
\ee
we obtain the following asymptotic expansion for $P(\r)$ \cite{HoyosBadajoz:2008fw}:
\be
P(\r)=c e^{4\rho/3}+\frac{4N_c^2}{c}\left( \r^2-\r+\frac{13}{16}\right)e^{-4\rho/3}+\co(\r e^{-8\rho/3}).
\ee
Via (\ref{changevariables}) then we obtain 
\bea
e^{2q}&=&\frac{c}{4}e^{4\r/3}+\frac{N_c}{4} (2 \r-1)+\co(e^{-4\r/3}),\NO\\
e^{2g}&=&ce^{4\r/3}+N_c(1-2\r)+\co(e^{-4\r/3}),\NO\\
e^{2k}&=&\frac{2 c}{3 }e^{4\r/3}-\frac{N_c^2}{6c}(4\r-5)^2e^{-4\r/3}+\co(e^{-8\r/3}),\NO\\
e^{4(\F-\F_o)}&=&\frac{3}{2c^3}\left(1+\frac{3N_c^2}{4c^2}(1-8\r)e^{-8\r/3}+\co(e^{-4\r})\right),\NO\\
a&=&2e^{-2\r}-\frac{2N_c}{c}(1-2\r)e^{-10\r/3}+\co(e^{-14\r/3}).
\eea

\begin{flushleft}
{\em IR asymptotics}
\end{flushleft}

As $t\to 0$ we have
\be
g_0(t)=\frac{N_c^2}{9}\left((3t)^{4/3}-\frac25(3t)^2+\co(t^{8/3}\right).
\ee
With the choice $t_{2k}^-=0$ discussed above then, it is easy to show that 
$p_{2k}(t)$ admits an expansion of the form
\be
p_{2k}(t)=\sum_{\ell=0}^\infty p_{2k}^{(\ell)}t^{\frac{2\ell+4}{3}},
\ee   
where
\be
p_{0}^{(0)}=1, \quad p_{0}^{(\ell)}=0,\; \ell>0,\quad p_{2k}^{(0)}=-\int_0^\infty dt't'^{-7/3}R_{2k}(t'),\; k>0.
\ee
Hence,
\be\label{unflavored-IR-expansion}
\cp(t)=\sum_{k=0}^\infty \tilde{c}^{1-2k}\sum_{\ell=0}^\infty p_{2k}^{(\ell)}t^{\frac{2\ell+4}{3}}
=:\sum_{\ell=0}^\infty p^{(\ell)}(\tilde{c})t^{\frac{2\ell+4}{3}}.
\ee
In principle, one can obtain this expansion directly from the recursion relations (\ref{c-expansion})
as we did for the UV asymptotics, but each term in the IR expansion will contain an infinite sum over
powers of $\tilde{c}$:
\be
p^{(\ell)}=\sum_{k=0}^\infty \tilde{c}^{1-2k}p_{2k}^{(\ell)}.
\ee 
Crucially, all these sums can be expressed algebraically in terms of the leading term $\tilde{h}_1\equiv p^{(0)}$. One
then obtains the expansion 
\be
\cp=\tilde{h}_1t^{4/3}-\frac{9N_c^23^{-2/3}}{40 \tilde{h}_1}t^2
-\frac{9N_c^2\left(81N_c^23^{-4/3}-160\tilde{h}_1^2\right)}{44800\tilde{h}_1^3}t^{8/3}+\co(t^{10/3}).
\ee
Alternatively, this expansion can be obtained by inserting (\ref{unflavored-IR-expansion}) in
(\ref{master-simple}). Differentiating this with respect to $t$ now and using
\be
t=\frac{8 \rho ^3}{3}+\frac{32 \rho ^5}{15}+\frac{256 \rho ^7}{315}+\co(\r^9),
\ee
we obtain the IR expansion \cite{Casero:2006pt,HoyosBadajoz:2008fw}
\be
P= h_1 \r+ \frac{4 h_1}{15}\left(1-\frac{4 N_c^2}{h_1^2}\right)\r^3
+\frac{16 h_1}{525}\left(1-\frac{4N_c^2}{3h_1^2}-\frac{32N_c^4}{3h_1^4}\right)\r^5+\co(\r^7),
\ee
where we have defined
\be
\tilde{h}_1=\frac{3^{4/3}}{8}h_1.
\ee
Using this expansion then we get via (\ref{changevariables})
\bea
e^{2q}&=&\frac{h_1 \rho ^2}{2}+\frac{4}{45} \left(-6 h_1+15 N_c-\frac{16 N_c^2}{h_1}\right) \rho ^4+\co(\r^6),\NO\\
e^{2g}&=&\frac{h_1}{2}+\frac{4}{15} \left(3 h_1-5 N_c-\frac{2 N_c^2}{h_1}\right) \rho ^2+\frac{8 \left(3 h_1^4+70 h_1^3 N_c-144 h_1^2 N_c^2-32 N_c^4\right) \rho ^4}{1575 h_1^3}\NO\\
&&+\co(\r^6),\NO\\
e^{2k}&=&\frac{h_1}{2}+\frac{2\left(h_1^2-4 N_c^2\right) \rho ^2}{5 h_1}+\frac{8\left(3 h_1^4-4 h_1^2 N_c^2-32 N_c^4\right) \rho ^4}{315 h_1^3}+\co(\r^6),\NO\\
e^{4(\F-\F_o)}&=&\frac{32}{h_1^3}\left(1+\frac{64  N_c^2 \rho ^2}{9 h_1^2}+\frac{128 N_c^2 \left(-15 h_1^2+124 N_c^2\right) \rho ^4}{405 h_1^4}+\co(\r^6)\right),\\
a &=&1+\left(-2+\frac{8 N_c}{3 h_1}\right) \rho ^2+\frac{2 \left(75 h_1^3-232 h_1^2 N_c+160 h_1 N_c^2+64 N_c^3\right) \rho ^4}{45 h_1^3}+\co(\r^6).\NO
\eea

\begin{figure}[ht]
\begin{minipage}[l]{0.5\linewidth}
\centering
\includegraphics[scale=.8]{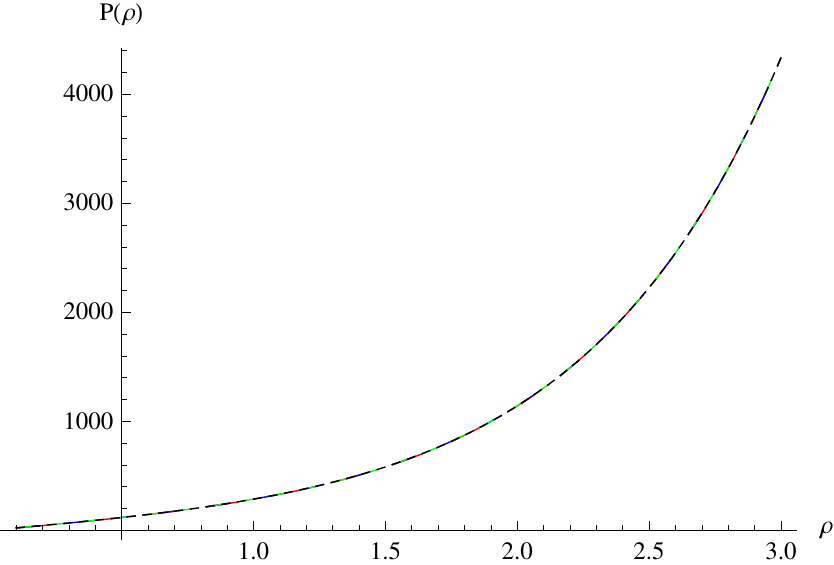}
\end{minipage}
\hspace{0.5cm}
\begin{minipage}[l]{0.5\linewidth}
\centering
\includegraphics[scale=.8]{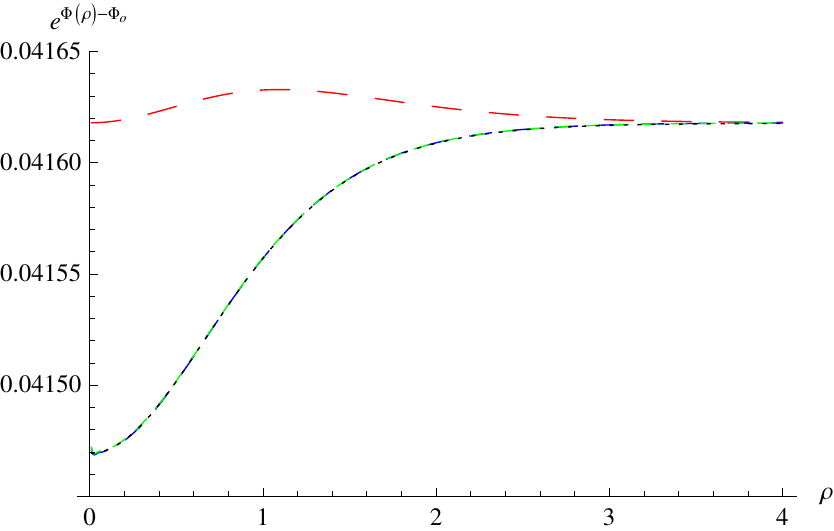}
\end{minipage}
\caption{The function $P(\r)$ and the dilaton  plotted using the first three orders in the expansion 
(\ref{c-expansion}) ($\co(c)$ red, $\co(c^{-1})$ blue, and $\co(c^{-3})$ green) are compared to the numerical solution. 
The plots are for the values $c=79.370$, $N_c=10$ and $h_1=223.3$, the last of which is determined numerically. Note
that the numerical value for $h_1$ is very close to the leading order approximation $h_1\approx 4 c/3^{1/3}$ in the
expansion (\ref{UV-IR-relation}) relation, which, as seen in the plot of the dilaton, converges very fast.}
\label{plots-unflavored}
\end{figure}

Before we move on to the flavoured solution, let us point out a couple of interesting properties of the 
unflavoured solution, which will not be shared with its flavoured version. Firstly, the fact that the 
expansion (\ref{c-expansion}) is uniformly valid for all $t$, implies that not only both the UV and IR asymptotic
expansions can be directly obtained from the large-$c$ expansion as we discussed above, but we can also 
relate the IR parameter $h_1$ with the UV parameter $c$. Namely, we have
\be\label{UV-IR-relation}
h_1=\frac{8}{3^{4/3}}\left(\frac{3c}{2}-\sum_{k=1}^\infty \left(\frac{3c}{2}\right)^{1-2k}
\int_0^\infty dt t^{-7/3}R_{2k}(t)\right),
\ee
which provides a systematic way to obtain $h_1$ in an expansion for large $c$ to any desired order. Since 
the dilaton goes to a constant both in the UV and in the IR, as seen in the above asymptotic expansions, we
can view (\ref{UV-IR-relation}) as a relation between the UV and IR values of the dilaton. In fact, for $N_f=0$,
it follows from (\ref{BPS-sols}) and (\ref{master-transformed}) that the Hamiltonian (\ref{effective-hamiltonian})
is essentially the dilaton \cite{Maldacena:2009mw}:
\be
H=4e^{-2(\F-\F_o)},
\ee
and hence (\ref{UV-IR-relation}) relates the IR and UV values of the Hamiltonian. As can be seen in Fig. \ref{plots-unflavored},
the expansion (\ref{c-expansion}) and the resulting relation (\ref{UV-IR-relation}) converge very fast.

\subsection{The flavoured solution}

Let us now try to construct the flavoured solution of (\ref{master-simple}) that reduces to the unflavoured
solution presented above as $N_f\to 0$. As we shall see, in the flavoured case the expansion (\ref{c-expansion}) is
not as useful as in the unflavoured case since it breaks down in the IR and hence it is not uniformly valid for all $t$. It nevertheless provides a useful representation of the solution in the UV.    

\begin{flushleft}
{\em UV asymptotics}
\end{flushleft}

The derivation of the UV asymptotics from (\ref{c-expansion}) proceeds as in the unflavoured case. As $t\to \infty$ we have
\bea
&&g_0(t)=\frac{1}{16}(2N_c-N_f)^2\left(\log^2(8t)-4\log(8t)+4+\co\left(\frac{\log^2t}{t}\right)\right),\NO\\
&&g_1(t)=\frac{N_f}{t}\left(1+\co\left(\frac{\log t}{t}\right)\right).
\eea
Using the recursion relations (\ref{large-c}), we compute
\bea
&&p_1(t)=\frac{9N_f}{8}t+\co(\log t),\\
&&p_2(t)=\frac{9}{8}t^{2/3}\left(\frac{1}{16}(2N_c-N_f)^2\left(\log^2(8t)-7\log(8t)+\frac{47}{2}\right)
-\left(\frac{9N_f}{8}\right)^2\right)+\co\left(\frac{\log^2(t)}{t^{1/3}}\right).\NO
\eea
Differentiating these with respect to $t$ we now obtain the asymptotic expansion \cite{HoyosBadajoz:2008fw}:
\be
P(\r)=c e^{4\r/3}+\frac{9N_f}{8}+\frac{1}{c}\left((2N_c-N_f)^2
\left(\r^2-\r+\frac{13}{16}\right)-\frac{81N_f^2}{64}\right)e^{-4\r/3} +\co(\r e^{-8\rho/3}).
\ee 
Via (\ref{changevariables}) these lead to
\bea
e^{2q}&=&\frac{c}{4}e^{4\r/3}+\frac{1}{8} \left((2 N_c-N_f)(2 \r-1)+\frac{9 N_f}{4}\right)+\co(e^{-4\r/3}),\NO\\
e^{2g}&=&ce^{4\r/3}+\frac12\left((2 N_c-N_f)(1-2\r)+\frac{9N_f}{4}\right)+\co(e^{-4\r/3}),\NO\\
e^{2k}&=&\frac{2 c}{3 }e^{4\r/3}+\frac{N_f}{2}-\frac{1}{24c}\left((2N_c-N_f)^2(4\r-5)^2-\left(\frac{9N_f}{2}\right)^2\right)
e^{-4\r/3}+\co(e^{-8\r/3}),\NO\\
e^{4(\Phi-\Phi_o)}&=&\frac{3}{2 c^3}\left(1-\frac{3N_f}{c}e^{-4\r/3}+\frac{3}{16c^2}\left((2N_c-N_f)^2(1-8\r)+297N_f^2\right)
e^{-8\r/3}+\co(e^{-4\r})\right),\NO\\
a&=&2e^{-2\r}-\frac{1}{c}(2N_c-N_f)(1-2\r)e^{-10\r/3}+\co(e^{-14\r/3}).
\eea

\begin{flushleft}
{\em IR asymptotics}
\end{flushleft}

Turning to the IR asymptotics, we need the expansions
\bea
&&g_0(t)=\left(\frac{2N_c-N_f}{2}\right)^2\frac19\left((3t)^{4/3}-\frac25(3t)^2+\co(t^{8/3})\right),\NO\\
&&g_1(t)=2N_f\left((3t)^{-2/3}-\frac15+\co(t^{2/3})\right).
\eea
Inserting these in the recursion relations (\ref{c-expansion}) we find for small $t$
\be\label{IR-solution-1}
\cp(t)=\tilde{h}_1 t^{4/3}-\frac{N_f}{3^{2/3}2}\left(1-\frac{N_f}{3^{2/3}8 \tilde{h}_1}
+\frac{N_f^2}{3^{1/3}32\tilde{h}_1^2}\right)t^{4/3}\log t
+\frac{N_f^3}{576\tilde{h}_1^2}t^{4/3}\log^2t+\ldots.
\ee
It can be shown that this expansion is what one obtains by first solving (\ref{master-simple}) perturbatively
in $N_f$ around the unflavoured solution of the previous section and {\em 
then} looking at the IR behaviour of this
solution. However, it is obvious that this expansion is not valid all the way down to $t=0$ as it breaks
down at $t\approx \exp(-3^{2/3}2\tilde{h}_1/N_f)$. It follows that the expansion (\ref{c-expansion}) breaks 
down in the IR for $N_f\neq 0$ and (\ref{IR-solution-1}) is not the correct IR 
behaviour. To recover 
the correct IR behaviour of the flavoured solution we need to go back to 
(\ref{master-simple}) and look
for an asymptotic expansion as $t\to 0$ {\em non-perturbatively} in $N_f$.  

It can be shown that such an asymptotic solution of (\ref{master-simple}) is given by
\bea\label{IR-solution-2}
\cp(t)&=&\frac{N_f}{3^{2/3}2}t^{4/3}\left(-\log t+\frac{3^{2/3}2}{N_f}\tilde{h}_1
-\frac14\log (-\log t)-\frac{1}{16}\frac{\log (-\log t)}{\log t}\right.\NO\\
&&\left.-\frac14\left(\frac{3^{2/3}2}{N_f}\tilde{h}_1-\frac54\right)\frac{1}{\log t}
+\co\left(\frac{\log (-\log t)}{(\log t)^2}\right)\right),
\eea
where again $\tilde{h}_1$ is an arbitrary constant. Differentiating with respect to $t$ and using 
\be
t=\frac{8 \rho ^3}{3}+\frac{32 \rho ^5}{15}+\frac{256 \rho ^7}{315}+\co(\r^9),
\ee
we obtain 
\be
P(\r)=h_1\r+\frac{4N_f}{3}\left(-\r\log\r-\frac{1}{12}\r\log(-\log\r)+\co\left(\frac{\r\log (-\log \r)}{\log \r}\right)\right)
+\co(\r^3\log \r),
\ee
which leads via (\ref{BPS-sols}) to
\bea
e^{2q}&=&\frac12\left(-\frac{4N_f}{3}\log\r+ h_1-\frac{N_f}{9}\log(-\log\r)+\co\left(\frac{\log (-\log \r)}{\log \r}\right)\right)\r^2+\co(\r^3\log\r),\NO\\
e^{2g}&=&\frac12\left(-\frac{4N_f}{3}\log\r+ h_1-\frac{N_f}{9}\log(-\log\r)+\co\left(\frac{\log (-\log \r)}{\log \r}\right)\right)+\co(\r\log\r),\NO\\
e^{2k}&=&\frac12\left(-\frac{4N_f}{3}\log\r+ h_1-\frac{N_f}{3}-\frac{N_f}{9}\log(-\log\r)+\co\left(\frac{\log (-\log \r)}{\log \r}\right)\right)+\co(\r^2\log\r),\NO\\
e^{4(\F-\F_o)}&=&\frac{27}{2N_f^3(-\log\r)^3}\left(1-\frac{\log(-\log\r)}{4\log\r}
+\co\left(\frac{1}{\log\r}\right)\right),\NO\\
a&=&1-2\r^2\left(1+\frac{(2N_c-N_f)}{2N_f\log\r}+\co\left(\frac{\log(-\log\r)}{(\log\r)^2}\right)\right)+\co(\r^3\log\r).
\eea
Note that, contrary to the flavour-less case, the dilaton goes to $-\infty$ in the IR and not to a constant. Moreover, the
Hamiltonian (\ref{effective-hamiltonian}) for $N_f\neq 0$ is not simply related to the dilaton.

\section{An analytic solution for $N_f=2N_c$}
\label{ananal}

Here we summarise some results about an analytic 
solution that is obtained as the rotation from the one 
with $N_f=2N_c$. 
We will use as ``seed'' solution the solution quoted in eqs.(4.7)-(4.8) of 
the paper \cite{HoyosBadajoz:2008fw}. Notice that these solutions are 
qualitatively different from those
studied in the main text, for example, 
the fibrations in the metric are absent, and the function 
$Q(\r)$ is constant. The functions $P(\r),Q(\r)$ are  
known exactly in this 
case and given by
\be
P=\frac{9 N_c}{4}+ c e^{\frac{4}{3}\r},\;\;\; Q= \pm \frac{3 
N_c}{4},\;\;\; \cosh\tau=1.
\ee
We will keep only the upper sign in the following. In this case the radial 
coordinate ranges in the whole real axis. For large 
values of $\rho$ the geometry asymptotes to the conifold, while in the 
IR, for $\r\to -\infty$, the geometry is the one for $N_f=2N_c$ with a 
linear dilaton, obtained in eqs.(4.22)-(4.23) of the paper 
\cite{Casero:2006pt} (for the value of $3\xi=4$).
Using (\ref{dildil}), we find functions of the background read
\bea
\label{newexactnf=2nc}
e^{2q}=\frac{3N_c}{4}+\frac{c}{4}e^{4\r/3}, \quad 
\frac{e^{2g}}{4}=\frac{3N_c}{8}+\frac{c}{4}e^{4\r/3}, 
\quad \frac{e^{2k}}{4}=\frac{N_c}{4}+\frac{c}{6}e^{4\r/3},\nn\\
\quad e^{-4\f}=e^{-4\f_{\infty}}
[1+\frac{3N_c e^{-4\r/3}}{2c}][1+\frac{9N_c e^{-4\r/3}}{2c} 
+\frac{9N_c^2 e^{-8\r/3 }}{2c^2}],\nn\\
\quad \hat{h}= e^{-2\f} - e^{-2\f_{\infty}}, ~~~~~a=b=0.
\label{zznf=2nc}
\eea
The warp factor  $\hat{h}$ for large values of the radial  coordinate ($\rho \to \infty$) has the following expansion:
\be
\hat{h} = e^{-2\f_{\infty}}\frac{3N_c}{c}\Big[e^{-4\r/3} 
+\frac{3N_c}{8c}e^{-8\r/3} -\frac{27N_c^3}{128 c^3} e^{-16\r/3} 
+O(e^{-20\r/3})    \Big].
\label{warpfactornf2nc}
\ee
Asymptotically in $\rho \to\infty$, we have that the dilaton is constant
and the metric and fluxes read
\bea
&& \dd s_{10}^2\approx \sqrt{3c N_c} e^{2\r/3  +\f_\infty}\Big[ 
\frac{\dd x_{1,3}^2}{3N_c} + \frac{2}{3}\dd\r^2  +\frac{1}{4} (\dd \Omega_2 +\dd\tilde{\Omega}_2) + \frac{1}{6}(\dd\psi+\cos\theta \dd\varphi +\cos\tilde{\theta} 
\dd\tilde{\varphi})^2 
\Big]\nonumber\\
& & F_3 \approx -\frac{N_c}{4}\Big[\sin\theta \dd\theta \wedge \dd\varphi +
\sin\tilde{\theta} \dd\tilde{\theta}\wedge \dd\tilde{\varphi}  \Big]\wedge
(\dd\psi + \cos\theta \dd\varphi + \cos 
\tilde{\theta}d\tilde{\varphi}).\nonumber\\
& & B_3\approx {}^{*}_6 F_3,\;\; ~~~~
F_5 \approx(1+*_{10})\frac{4ce^{2\f_\infty}}{3N_c}e^{4\r/3}\vol_4 \wedge 
\dd\r.
\eea
On the other hand, when we explore the geometry near the IR ($\rho \to 
-\infty$ where there 
is a singularity), we will find that the whole background resembles that 
in eqs.(4.22)-(4.23) of the paper
\cite{Casero:2006pt} for the value of $3\xi=4$. The reason for this is 
that when going to the far IR, the rotation is ``undone'', 
 according to eq.(\ref{importanteq}) and the dilaton vanishing rapidly 
as $e^{4\f}|_{\r\to-\infty}\sim e^{4\r}$.

Notice that contrary to the $N_f=2N_c$ solution discussed in the main text,
this solution preserves the $U(1)_R$ symmetry, while it does not preserve any $\Z_2$
symmetry discussed in section \ref{thez2}. Further studies of these special solutions, and their field theory 
interpretations, are left for future work. 


\section{Flavoured fivebrane solution in the variables of  \cite{Maldacena:2009mw}}
\label{mmvar}

For convenience of the reader,  in this appendix we write the fivebrane solution
in the variables used in \cite{Maldacena:2009mw}. In particular, the complete solution 
may be written in terms of two functions\footnote{The function $c$ in \cite{Maldacena:2009mw} is not to be confused with the parameter $c$ used elsewhere in this paper.} $c, f$, obeying two coupled first order 
differential equations.  The variable $t$ used here is related to that used in the main text as 
$t=2\rho$. Following  \cite{Maldacena:2009mw} we write the solution for NS 
fivebranes, with the metric in string frame: 
\bea 
\label{ansatz}
\dd s^2_{str} &=& \dd x_{3+1}^2 + { N_c\over 4 }  \dd s^2_6
\\[2mm]
\dd s^2_6 &=&  \left(c' +\frac{N_f}{2N_c}\right) ( \dd t^2 + (\epsilon_3 + A_3)^2) +
{ c \over \tanh t} ( \epsilon_1^2 + \epsilon_2^2 + e_1^2 + e_2^2)  +
2 { c \over \sinh t} ( \epsilon_1 e_1 + \epsilon_2 e_2)  \notag
\\[2mm] &&  - \left(1 - \frac{N_f}{2N_c} \right)  \left( { t \over \tanh t } - 1 \right)
( \epsilon_1^2 + \epsilon_2^2 - e_1^2 - e_2^2)\label{themetric}
\\[2mm]
e^{ 2 \Phi} &=& e^{ 2 \Phi_0} { f^{1/2} \over \sinh^2 t  } \left(c' +\frac{N_f}{2N_c}\right)  \label{dilatonvalue}
\\[2mm]
\frac{4}{N_c} H_3 &=&   ( \epsilon_3 + A_3)\wedge \left[  \epsilon_1 
\wedge \epsilon_2  +\left(1-\frac{N_f}{N_c}\right) e_1 \wedge e_2  +\left(1 - \frac{N_f}{2N_c} \right) {t \over \sinh t}
( \epsilon_1 \wedge e_2 + e_1 \wedge  \epsilon_2 )\right]  \nonumber \\
&&  ~~~~ +\left(1 - \frac{N_f}{2N_c} \right) {(t \coth t -1 )\over \sinh t}
\dd t \wedge ( \epsilon_1 \wedge e_1 + \epsilon_2 \wedge e_2 )
\eea
 where
 \bea
 e_1 &=& \dd  \theta_1 ~,~~~~~~e_2 = - \sin \theta_1 \dd\phi_1 ~,~~~~~~~~~~A_3 = \cos \theta_1 \dd\phi_1~,\notag
 \\
 \epsilon_1 + i \epsilon_2 &=& e^{ - i \psi} ( \dd\theta_2 + i \sin \theta_2 \dd \phi_2 ) ~,~~~~~~~
 \epsilon_3 = \dd \psi + \cos \theta_2 \dd \phi_2~.
 \eea
The $SU(2)$ left-invariant one-forms $\epsilon_i$ obey $\dd \epsilon_1 = - \epsilon_2 \wedge \epsilon_3$ and cyclic permutations.
The functions $c(t)$ and $f(t)$ appearing in (\ref{ansatz}) obey the equations
 \bea
 f ' &=& 4 \sinh^2 t \, c \label{equationf}
 \\
 c' &=& { 1 \over f } [ c^2 \sinh^2 t - (t \cosh t -  \sinh t )^2 ]   - \frac{N_f}{2N_c}
\label{equationc}
 \eea
where the primes denote derivatives with respect to $t$.  This system 
is equivalent to the second order equation
\bea
4 f f'' - f'(f' + 8 f \coth t) + 16 \sinh^2 t \left[ \left(1-\frac{N_f}{2N_c}\right)^2 (\sinh t - t \cosh t)^2  + \frac{N_f}{2N_c}
 f  \right] = 0 ~.~~~
\label{masterised}
\eea
Setting $N_f=0$, the ansatz and the differential equations reduce exactly 
to those in  \cite{Maldacena:2009mw}. 

Notice the three-form can be written as 
\bea
H_3 &=&   - \frac{N_f}{4}( \epsilon_3 + A_3)\wedge e_1 \wedge e_2  + \mathrm{closed}\label{close}
\eea
In these variables it is simple to study the $\Z_2$ action generated by \cite{Klebanov:1998hh}
\bea
{\cal I}: ~~\theta_1 \leftrightarrow \theta_2, ~~~~ \phi_1 \leftrightarrow \phi_2, ~~~~ H_3 \to -H_3, ~~~~ F_3 \to - F_3, 
\eea
where the action on the three-forms is generated by the center of the $SL(2,\Z)$ action. When $N_f=0$, the three-form $H_3$
is invariant under ${\cal I}$, while the metric is not invariant. Therefore the full solution breaks this $\Z_2$ symmetry \cite{Butti:2004pk}.
On the other hand, when $N_f$ is non-zero, $H_3$ is not invariant under ${\cal I}$. In particular, the non-closed part of $H_3$
in  (\ref{close}) is not invariant, while the closed part is still invariant. In the particular case that $N_f=2N_c$ the metric
becomes manifestly invariant under ${\cal I}$, however the non-closed part of $H_3$ is not invariant. In this case
the three-form becomes simply
\bea
H_3 &=&    \frac{N_c}{4}( \epsilon_3 + A_3)\wedge\left[ \epsilon_1 \wedge \epsilon_2 +  e_1 \wedge e_2  \right].
\label{simpleh}
\eea
This is  manifestly not invariant under ${\cal I}$, 
however it is clearly invariant under the swap of two-spheres, without 
change of sign of $H_3$. Therefore, for $N_f=2N_c$ 
the full solution is invariant under  $\theta_1 \leftrightarrow \theta_2,  \phi_1 \leftrightarrow \phi_2$.

\section{More on the flavoured quiver}
\label{quiverappendix}

In this appendix we make  some comments on the quiver gauge theory 
discussed  in section \ref{sce1} of the main body of 
the paper, see  eq.(\ref{newprop}). 
In particular, here we will study Seiberg dualities and beta functions.
The quiver may be written as 
\be
SU(N_1) \times SU(N_2)\times  SU(N_f/2)_\mathrm{flavour}
\ee
and is described in Figure \ref{shittyquiver} in the main text.

\subsection{Going to the IR} 

We start performing Seiberg dualities. After 
$z$ Seiberg  dualities, the quiver will look like
\be
 SU\Big[\frac{z(z-1)N_f}{4} + z N_2  +(1-z)N_1\Big] 
\times
SU\Big[\frac{z(z+1)N_f}{4} - z N_1 +(1+z)N_2\Big]\times 
SU(N_f/2)_\mathrm{flavour}
\ee
where the flavour group is untouched.
Assuming that the anomalous dimensions of the $A_i, B_i$ fields are 
$\gamma_{A,B}=-\frac{1}{2}$,  (with no assumptions  made on the dimensions 
of the quark superfields),  the beta functions of each group, at each step of the cascade are
 \bea
& &\beta_1=(-1)^z 3[N_1-N_2 - z\frac{N_f}{2}] - \frac{N_f}{2}(1-\gamma_q)   
\nonumber\\
& &\beta_{2}=(-1)^{z+1} 3[N_1-N_2 - z\frac{N_f}{2}] - 
\frac{N_f}{2}(1-\gamma_q)
\eea
and the change of each group in a given step of the 
cascade is\footnote{The change of the group is defined as 
Rank[before]-Rank[after], and is 
typically positive indicating a decrease of 
the group and degrees of 
freedom under  Seiberg dualities down the cascade. },
\bea
& &\Delta N_{1} =\frac{1+(-1)^{z+1}}{2}[2 N_1-2 N_2 - \frac{N_f}{2}(2z-1)]  
\nonumber\\
& & \Delta N_{2} =\frac{1+(-1)^{z}}{2}[2 N_1-2 N_2 - \frac{N_f}{2}(2z-1)].
\eea
Notice that for odd steps, the group 2
does not change while for even one the group 1 does not change. 
If we use as a criteria for the existence of a cascade down 
the flow, the fact that the changes in groups must be positive, whenever we find a 
negative $\Delta N_i$ means  that we must  stop the duality cascade.
This bounds the number of dualities performed, by a critical number given by
\be
z_*=\mathrm{Int}[\frac{1}{2} +\frac{2(N_1-N_2)}{N_f}]
\ee
where with $\mathrm{Int}$ we mean the integer part.
Following the discussion in the main part of the paper, for simplicity
let us  consider the case when  
\be
N_1=[k+1]N_c, \;\;\;\; N_2=k N_c, \;\;\;\; N_f= \frac{4N_c}{\lambda},\;\; 
z_*= \mathrm{Int} \Big[\frac{\lambda +1}{2}\Big].
\label{kkktmba}
\ee
Notice that for $0<\lambda<\infty$, we can have $N_f$ small 
or large compared to $N_{1},N_2$. 
Also, notice that $\lambda$ must be a rational number.
For example, if we restrict the attention in the
 interval $N_c<N_f<2 N_c$,
we have $2<\lambda<4$. This in turn bounds the maximum 
number of Seiberg dualities to $z_*<3$. It may be instructive to consider
the cases $\lambda=2,3,4$, to  notice that the pattern is that after one or two dualities, the quiver 
comes back to itself, or is in a distribution of colours and flavours that 
does not permit further Seiberg dualities.  Although from the field theory point of view, we could 
consider the situation in which $\lambda \gg 1$, in which we can do many Seiberg dualities, 
our string theory backgrounds certainly do not require this. In fact, we typically have $\lambda \sim {\cal O}(1)$.
Notice that, on the contrary, 
in the situation studied in \cite{Benini:2007gx}
we have that $\frac{N_f}{N_1-N_2}\to 0$. 
This gives the  possibility of performing
many Seiberg dualities, and in fact  $z_{*}$ diverges  there.

\subsection{Going to the UV}

We will quote here similar formulas to the ones above, for the flow 
towards the UV.
As usual, the idea is to  Seiberg dualise the strongly coupled group.
The first point is that after $s$ Seiberg dualities (we distinguish with 
$s$ the dualities to the UV while with $z$ those to the IR), we have a 
quiver
\be
 SU\Big[\frac{s(s-1)N_f}{4} + s N_1 
+(1-s)N_2\Big] 
\times
SU\Big[\frac{s(s+1)N_f}{4} - s N_2 +(1+s)N_1 \Big]\times 
SU(N_f/2)_\mathrm{flavour}
\ee
The beta functions at each step  are
\bea
\beta_1 &=&(-1)^{s+1} 3[N_2-N_1 - s\frac{N_f}{2}] - 
\frac{N_f}{2}(1-\gamma_q),\\
\beta_{2}&=&(-1)^{s} 3[N_2-N_1 - s\frac{N_f}{2}] - 
\frac{N_f}{2}(1-\gamma_q).
\eea
Notice that the beta function for the sum of the gauge couplings is 
negative 
(assuming $\gamma_q<1$), independently of the number of Seiberg dualities. This implies that at some point in the UV at least 
one of the couplings will diverge, that is, there is a Landau pole. This behaviour was indeed observed in the solution of \cite{Benini:2007gx}, which was
proposed to be dual to the quiver we are discussing, but 
at the origin of moduli space, \emph{i.e.} not in a Higgsed phase.

The change in groups is
\bea
\Delta N_{1} &=&\frac{1+(-1)^{s}}{2}[2 N_2-2 N_1 - 
\frac{N_f}{2}(2s-1)], \\
\Delta N_{2} &=&\frac{1+(-1)^{s+1}}{2}[2 N_2-2 
N_1 - \frac{N_f}{2}(2s-1)]
\eea
Now, going to the UV, one would in principle keep on doing Seiberg dualities  unless 
$\Delta N_i>0$, which would mean that there is a decrease in the number of 
degrees of freedom. This implies that
\be
\Delta N_i>0 \to s< \frac{1}{2}-\frac{2M}{N_f}.
\ee
Coming back to  eq.(\ref{kkktmba}), this implies that for $s< \frac{1-\lambda}{2}$
we have to stop dualising. For any $s>0$, we have that the  inequality below
cannot be satisfied if $M>0, N_f>0, s>1$, so, if we can do one Seiberg duality, 
we can do  as many as we want. As we can see,  
this quiver has non-stopping Seiberg dualities to the UV. 
In the case of \cite{Benini:2007gx}, one can perform many Seiberg dualities before 
reaching a ``duality wall''. 
This occurs at some finite energy scale, where the Seiberg 
dualities ``accumulate'', and the number of degrees of freedom 
(for example measured by the central charge) diverge.

\section{The holographic central charge}
\label{cchargeapp}

In this appendix we compute the  ``holographic central charge''  for our  backgrounds.  
To define this quantity, the idea is to reduce the ten-dimensional system to a 
five-dimensional sigma model coupled to scalars, following for example 
the procedure described in \cite{Berg:2005pd}. Then one uses results 
derived in \cite{Girardello:1998pd}, to show that such quantity is monotonic, and stationary at 
 AdS points. Thus it may be interpreted as a central function, measuring the degrees of freedom 
of the field theory. Defining the functions
\be
H=\frac{\hat{h}}{16}e^{6\phi+4h+4g+2k},\;\;\;~~~~ \beta=\hat{h}e^{2\phi+2k}
\ee
and reducing our background in eq.(\ref{rotated-background}) to
a five-dimensional gravity theory coupled to a sigma model of scalars, 
we  find (in the Einstein frame of the five-dimensional theory) that 
the metric reads
\be
\dd s_{5,E}^2= H^{1/3}[\dd x_{1,3}^2+ \beta \dd \rho^2].
\ee
One can define a quantity, that may be
identify with the central charge (more precisely, central function) as
\be
{\mathbf c}\propto \frac{ \beta^{3/2} H^{7/2}}{H'^3}.
\ee
Specifying this to our background we find
\be
{\mathbf c} \propto \hat{h}^2 e^{6\phi+4k+2h+2g}
\Big[\frac{ \hat{h}' }{\hat{h}} +(4g'+4h'+2k'+6\phi')  \Big]^{-3}.
\ee

We will evaluate this expression at large radius. 
Using the UV expansions
of the various functions, we see 
the term in square brackets is a numerical coefficient, 
which we will ignore.
The factors of $c$ and of $g_s$ 
cancel between the exponentials and the warp factor, and we pick up 
precisely the coefficient in the numerator of the warp factor in (\ref{truewarp}), namely we get
\bea
{\mathbf c} \sim (n_f+n)^2 ~~~\mathrm{for} ~~~~r\to \infty.
\eea

This is the expected behaviour for the theory 
$SU(N_c+n+n_f) \times SU(N_f/2+n+n_f)$, in the limit that $n+n_f\gg N_c$ and $n+n_f\gg N_f$.
Of course the number of D3 branes is running, and using this,  we have
\bea
{\mathbf c} \sim (\nu r^2+(N_c-N_f/2) ^2 \log r)^2 ~~~\mathrm{for} ~~~r\to \infty.
\eea
The result for the unflavoured Klebanov-Strassler theory   ($\nu=0$, $N_f=0$) 
 was obtained in \cite{Klebanov:2007ws}.
We see that in our case the dominating degrees of freedom in the UV are the $n_f$ 
source D3 branes. The logarithmic growth is related to the cascading behaviour, 
whereas the much more rapid power-law growth that we see is presumably due to the Higgsing. 

Note that  $H'$ does not vanish anywhere, and correspondingly the central function 
$\mathbf{c}$ is monotonically increasing up to the UV. In particular, 
our solution does not display
a duality wall, as  was observed in the solution 
 of \cite{Bigazzi:2009gu},  where a divergence of $\mathbf{c}$ at a finite value 
of the radial coordinate was shown to exist. 
 If  there is a Landau pole in the field theory,
it will occur at infinite energy, where the number of degrees of freedom
measured by the quantity $\mathbf{c}$ defined above,  also diverges.


\begin{thebibliography}{99}

\bibitem{Butti:2004pk}
  A.~Butti, M.~Grana, R.~Minasian, M.~Petrini and A.~Zaffaroni,
  ``The baryonic branch of Klebanov-Strassler solution: A supersymmetric
  family of $SU(3)$ structure backgrounds,''
  JHEP {\bf 0503}, 069 (2005)
  [arXiv:hep-th/0412187].

\bibitem{Grana:2005sn}
  M.~Grana, R.~Minasian, M.~Petrini and A.~Tomasiello,
  ``Generalized structures of ${\cal N}=1$ vacua,''
  JHEP {\bf 0511}, 020 (2005)
  [arXiv:hep-th/0505212].

\bibitem{Strominger:1986uh}
  A.~Strominger,
  ``Superstrings with Torsion,''
  Nucl.\ Phys.\  B {\bf 274}, 253 (1986).

\bibitem{Dasgupta:1999ss}
  K.~Dasgupta, G.~Rajesh and S.~Sethi,
  ``M theory, orientifolds and G-flux,''
  JHEP {\bf 9908}, 023 (1999)
  [arXiv:hep-th/9908088].

\bibitem{Maldacena:2009mw}
  J.~Maldacena and D.~Martelli,
  ``The unwarped, resolved, deformed conifold: fivebranes and the baryonic
  branch of the Klebanov-Strassler theory,''
  JHEP {\bf 1001}, 104 (2010)
  [arXiv:0906.0591 [hep-th]].

\bibitem{Minasian:2009rn}
  R.~Minasian, M.~Petrini and A.~Zaffaroni,
  ``New families of interpolating type IIB backgrounds,''
  arXiv:0907.5147 [hep-th].

\bibitem{Casero:2006pt}
  R.~Casero, C.~Nunez and A.~Paredes,
  ``Towards the string dual of ${\cal N} = 1$ SQCD-like theories,''
  Phys.\ Rev.\  D {\bf 73}, 086005 (2006)
  [arXiv:hep-th/0602027].

\bibitem{Nunez:2010sf}
  C.~Nunez, A.~Paredes and A.~V.~Ramallo,
  ``Unquenched flavor in the gauge/gravity correspondence,''
  arXiv:1002.1088 [hep-th].

\bibitem{HoyosBadajoz:2008fw}
  C.~Hoyos-Badajoz, C.~Nunez and I.~Papadimitriou,
  ``Comments on the String dual to ${\cal N}=1$ SQCD,''
  Phys.\ Rev.\  D {\bf 78}, 086005 (2008)
  [arXiv:0807.3039 [hep-th]].

\bibitem{Casero:2007jj}
  R.~Casero, C.~Nunez and A.~Paredes,
  ``Elaborations on the String Dual to ${\cal N}=1$ SQCD,''
  Phys.\ Rev.\  D {\bf 77}, 046003 (2008)
  [arXiv:0709.3421 [hep-th]].

\bibitem{Klebanov:2000nc}
  I.~R.~Klebanov and A.~A.~Tseytlin,
  ``Gravity Duals of Supersymmetric $SU(N) \times SU(N+M)$ Gauge Theories,''
  Nucl.\ Phys.\  B {\bf 578}, 123 (2000)
  [arXiv:hep-th/0002159].

\bibitem{cgnr}
E.~Conde et al.~ In preparation.

\bibitem{Andrews:2006aw}
  R.~P.~Andrews and N.~Dorey,
  ``Deconstruction of the Maldacena-Nunez compactification,''
  Nucl.\ Phys.\  B {\bf 751}, 304 (2006)
  [arXiv:hep-th/0601098].

R.~P.~Andrews and N.~Dorey,
  ``Spherical deconstruction,''
  Phys.\ Lett.\  B {\bf 631}, 74 (2005)
  [arXiv:hep-th/0505107].
  
  
\bibitem{Klebanov:2000hb}
  I.~R.~Klebanov and M.~J.~Strassler,
  ``Supergravity and a confining gauge theory: Duality cascades and
  $\chi_{SB}$-resolution of naked singularities,''
  JHEP {\bf 0008}, 052 (2000)
  [arXiv:hep-th/0007191].

\bibitem{Aharony:2000pp}
  O.~Aharony,
  ``A note on the holographic interpretation of string theory backgrounds  with  varying flux,''
  JHEP {\bf 0103}, 012 (2001)
  [arXiv:hep-th/0101013].

\bibitem{Martucci:2005ht}
  L.~Martucci and P.~Smyth,
  ``Supersymmetric D-branes and calibrations on general ${\cal N }= 1$ backgrounds,''
  JHEP {\bf 0511}, 048 (2005)
  [arXiv:hep-th/0507099].

\bibitem{Martelli:2003ki}
  D.~Martelli and J.~Sparks,
  ``$G$-structures, fluxes and calibrations in M-theory,''
  Phys.\ Rev.\  D {\bf 68}, 085014 (2003)
  [arXiv:hep-th/0306225].

\bibitem{Frey:2003sd}
  A.~R.~Frey and M.~Grana,
  ``Type IIB solutions with interpolating supersymmetries,''
  Phys.\ Rev.\  D {\bf 68}, 106002 (2003)
  [arXiv:hep-th/0307142].

\bibitem{Hull:1986kz}
        C.~M.~Hull,
        ``Compactifications of the Heterotic Superstring,''
        Phys.\ Lett.\  B {\bf 178}, 357 (1986).

\bibitem{Gauntlett:2001ur}
  J.~P.~Gauntlett, N.~Kim, D.~Martelli and D.~Waldram,
  ``Fivebranes wrapped on SLAG three-cycles and related geometry,''
  JHEP {\bf 0111}, 018 (2001)
  [arXiv:hep-th/0110034].

\bibitem{GMW}
  J.~P.~Gauntlett, D.~Martelli and D.~Waldram,
  ``Superstrings with intrinsic torsion,''
  Phys.\ Rev.\  D {\bf 69}, 086002 (2004)
  [arXiv:hep-th/0302158].

\bibitem{Maldacena:2000yy}
  J.~M.~Maldacena and C.~Nunez,
  ``Towards the large $N$ limit of pure ${\cal N }= 1$ super Yang Mills,''
  Phys.\ Rev.\ Lett.\  {\bf 86}, 588 (2001)
  [arXiv:hep-th/0008001].
  A.~H.~Chamseddine and M.~S.~Volkov,
  ``Non-Abelian BPS monopoles in N = 4 gauged supergravity,''
  Phys.\ Rev.\ Lett.\  {\bf 79}, 3343 (1997)
  [arXiv:hep-th/9707176].

\bibitem{Gaillard:2008wt}
  J.~Gaillard and J.~Schmude,
  ``On the geometry of string duals with backreacting flavors,''
  JHEP {\bf 0901}, 079 (2009)
  [arXiv:0811.3646 [hep-th]].

\bibitem{Koerber:2007hd}
  P.~Koerber and D.~Tsimpis,
  ``Supersymmetric sources, integrability and generalized-structure
  compactifications,''
  JHEP {\bf 0708}, 082 (2007)
  [arXiv:0706.1244 [hep-th]].

\bibitem{Pando Zayas:2000sq}
  L.~A.~Pando Zayas and A.~A.~Tseytlin,
  ``3-branes on resolved conifold,''
  JHEP {\bf 0011}, 028 (2000)
  [arXiv:hep-th/0010088].

\bibitem{Martucci:2006ij}
  L.~Martucci,
  ``D-branes on general ${\cal N} = 1$ backgrounds: Superpotentials and D-terms,''
  JHEP {\bf 0606}, 033 (2006)
  [arXiv:hep-th/0602129].

\bibitem{Dymarsky:2005xt}
  A.~Dymarsky, I.~R.~Klebanov and N.~Seiberg,
  ``On the moduli space of the cascading $SU(M+p) \times  SU(p)$ gauge theory,''
  JHEP {\bf 0601}, 155 (2006)
  [arXiv:hep-th/0511254].

\bibitem{Krishnan:2008gx}
  C.~Krishnan and S.~Kuperstein,
  ``The Mesonic Branch of the Deformed Conifold,''
  JHEP {\bf 0805} (2008) 072
  [arXiv:0802.3674 [hep-th]].

\bibitem{Gubser:2004qj}
  S.~S.~Gubser, C.~P.~Herzog and I.~R.~Klebanov,
 ``Symmetry breaking and axionic strings in the warped deformed conifold,''
  JHEP {\bf 0409}, 036 (2004)
  [arXiv:hep-th/0405282].

\bibitem{Dymarsky:2009fj}
  A.~Dymarsky,
  ``Flavor brane on the baryonic branch of moduli space,''
  JHEP {\bf 1003}, 067 (2010)
  [arXiv:0909.3083 [hep-th]].

\bibitem{Benini:2007gx}
  F.~Benini, F.~Canoura, S.~Cremonesi, C.~Nunez and A.~V.~Ramallo,
  ``Backreacting Flavors in the Klebanov-Strassler Background,''
  JHEP {\bf 0709} (2007) 109
  [arXiv:0706.1238 [hep-th]].

\bibitem{Strassler:2005qs}
  M.~J.~Strassler,
  ``The duality cascade,''
  arXiv:hep-th/0505153.

\bibitem{Klebanov:1998hh}
  I.~R.~Klebanov and E.~Witten,
  ``Superconformal field theory on threebranes at a Calabi-Yau  singularity,''
  Nucl.\ Phys.\  B {\bf 536}, 199 (1998)
  [arXiv:hep-th/9807080].

\bibitem{Seiberg:1999vs}
  N.~Seiberg and E.~Witten,
  ``String theory and noncommutative geometry,''
  JHEP {\bf 9909}, 032 (1999)
  [arXiv:hep-th/9908142].

\bibitem{Bigazzi:2008qq}
  F.~Bigazzi, A.~L.~Cotrone, A.~Paredes and A.~V.~Ramallo,
  ``The Klebanov-Strassler model with massive dynamical flavors,''
  JHEP {\bf 0903}, 153 (2009)
  [arXiv:0812.3399 [hep-th]].

\bibitem{Berg:2005pd}
  M.~Berg, M.~Haack and W.~Mueck,
  ``Bulk dynamics in confining gauge theories,''
  Nucl.\ Phys.\  B {\bf 736}, 82 (2006)
  [arXiv:hep-th/0507285].

\bibitem{Girardello:1998pd}
  L.~Girardello, M.~Petrini, M.~Porrati and A.~Zaffaroni,
  ``Novel local CFT and exact results on perturbations of N = 4 super
  Yang-Mills from AdS dynamics,''
  JHEP {\bf 9812}, 022 (1998)
  [arXiv:hep-th/9810126].

  D.~Z.~Freedman, S.~S.~Gubser, K.~Pilch and N.~P.~Warner,
  ``Renormalization group flows from holography supersymmetry and a
  c-theorem,''
  Adv.\ Theor.\ Math.\ Phys.\  {\bf 3}, 363 (1999)
  [arXiv:hep-th/9904017].

\bibitem{Klebanov:2007ws}
  I.~R.~Klebanov, D.~Kutasov and A.~Murugan,
  ``Entanglement as a Probe of Confinement,''
  Nucl.\ Phys.\  B {\bf 796}, 274 (2008)
  [arXiv:0709.2140 [hep-th]].

\bibitem{Bigazzi:2009gu}
  F.~Bigazzi, A.~L.~Cotrone, A.~Paredes and A.~V.~Ramallo,
  ``Screening effects on meson masses from holography,''
  JHEP {\bf 0905}, 034 (2009)
  [arXiv:0903.4747 [hep-th]].


\end{thebibliography}
\end{document}